\documentclass[
reprint,
superscriptaddress,
frontmatterverbose, 
amsmath,
amssymb,
prapplied,
floatfix,
nofootinbib,
longbibliography
]{revtex4-2} 

\usepackage{graphicx}
\usepackage{dcolumn}
\usepackage{bm}
\usepackage{braket}
\usepackage{mathtools}
\usepackage{url}
\usepackage{dsfont}
\usepackage{xcolor}
\usepackage{float} 
\usepackage{lipsum}

\usepackage{pgfplots}
\pgfplotsset{compat=newest}
\usetikzlibrary{plotmarks}
\usetikzlibrary{arrows.meta}
\usepgfplotslibrary{patchplots}
\usepackage{grffile}

\begin{document}

\preprint{APS/123-QED}

\title{Active Acoustic Su–Schrieffer–Heeger-like Metamaterial}

\begin{abstract}
An acoustic dimer composed of two electronically controlled electro-acoustic 
resonators is presented in view of exploring one-dimensional topological phenomena. Active control allows for real-time manipulation of the metamaterial's properties, including its mechanical mass, resistance, compliance and internal coupling. The latter enables active tuning of the topological phase transition in the effective acoustic Su–Schrieffer–Heeger system. An analytical model of the unit cell as well as the electronic control scheme is shown. Band structures derived from the analytical model, the finite element simulation and the experimental data consistently demonstrate the realization of a tuneable one dimensional topological insulator.
\end{abstract}

\author{Mathieu~Padlewski}
\author{Maxime~Volery}
\affiliation{Signal Processing Laboratory 2, EPFL, 1015 Lausanne, Switzerland}
\author{Romain~Fleury}
\affiliation{Laboratory of Wave Engineering, EPFL, 1015 Lausanne, Switzerland}
\author{Hervé~Lissek}
\author{Xinxin~Guo}
\affiliation{Signal Processing Laboratory 2, EPFL, 1015 Lausanne, Switzerland}

\date{\today}

\maketitle


\section{Introduction}\label{sec:intr}

    The study of collective wave phenomena in condensed matter physics and photonics have proven to be an essential asset for the development of pioneering technologies that have shaped the modern world - transistors, photovoltaics, charge-coupled devices to name a few. A notable drawback to these fields of research, however, is the often highly complex and costly equipment required to observe microscopic phenomena occurring at the quantum scale. One solution would be to increase the scale of the physics being studied and to observe wave phenomena in macroscopic crystalline structures.
    
    Acoustic metamaterials (AMMs) are materials that can be engineered to exhibit properties similar to those found in quantum systems and even \textit{beyond} the realm of naturally occurring materials. In addition to their complexity and cost, macroscopic wave systems have notable advantages over those at the quantum scale. The salient physics of quantum systems, for example, is well defined in theory, but lacks atomic probes capable of readily determining eigenvalues and directly measuring eigenfunctions. On the other hand, classical analogues of these systems can offer complete observability and control of these quantities \cite{He1986,Maynard1992}. Generally speaking, AMMs can provide valuable insight into systems where theoretical calculations are not easily tractable, even with modern computers, particularly systems involving time-dependent potential fields, disorder, or non-linearity. 

    Topological insulators are a highly attractive area of research in condensed matter physics owing to features such as robust solitons or interface states. Unfortunately, progress in this field is heavily hindered by the limited number of available crystal candidates exhibiting such properties and the difficulty with which it is to carry out experiments. Conversely, heaps of analogous acoustic systems demonstrating topological regimes have been studied owing to the relative ease with which experimental fabrication and measurement are carried out \cite{Xiao2015,Zangeneh-Nejad2019,Yang2016, Esmann2018, Shen2020,Yan2020,Chen2020}. For example, Zhao \textit{et al.}\cite{Zhao2021} proposed an acoustic duct lined with Helmholtz resonators where a linear inter-site coupling is fixed by spatial separation between subsequent pairs in view of understanding topology within a local-resonant band gap. Effective controllability, tuneability and reconfigurability remain desirable properties that unfortunately reside beyond the scope of these passive systems.
    
    The emerging field of active acoustic systems on the other hand could provide phase transition and soliton formation manipulation through the application of active control schemes allowing for the exploration of new exotic phenomena. It's clear that advances in non-Hermitian PT-symmetric \cite{Auregan2017,Christensen2016,Fleury2015,Shi2016,Zhang2021,Liu2022}, dynamically reconfigurable \cite{Ma2016}, non-reciprocal systems \cite{Popa2014,Boechler2011,Zangeneh-nejad2018,Liang2010,Fleury2014,Fleury2015c} and real-time manipulation of the effective acoustic properties \cite{Popa2015,Cho2020} provide the necessary tools for seeking out new topological phenomena.

    The Active Electroacoustic Resonator (AER) allows achieving prescribed frequency-dependent acoustic impedances at the acoustic interface of an electroacoustic transducer such as a loudspeaker \cite{Rivet2017a}. Such devices have been shown to successfully achieve tunable sound absorption \cite{Rivet2017,Boulandet2018}. They have also been employed as active unit-cells in a non-Hermitian acoustic medium \cite{Rivet2018a}, and in an active acoustic metasurface concept for  shaping reflected wavefronts in a controllable manner \cite{Lissek2018}.

    This study presents an actively tuneable AMMs capable of exhibiting prescribed topological features similar to those found in the Su–Schrieffer–Heeger system \cite{Su1979} - a one dimensional chain of coupled dimers that can host topological modes owing to its two coupling-dependant regimes. Although successful realizations of this system have been previously achieved via passive meta-structures \cite{Coutant2021a,Zheng2019,Li2018}, here we demonstrate an actively tuneable dimer composed of two electronically-controlled AERs. First, an analytical model of the unit cell is presented followed by the active control scheme used to tune dimerisation of the unit cell. Fidelity of the model is demonstrated by comparing crystal dispersions obtained for both the model and the experimental unit cell. Using a transfer matrix approach, we finally demonstrate that the system can host topological edge states via a finite crystal simulation and that they are protected by both time-reversal symmetry and equality of the far-field scattering matrices.

\section{Description of the baseline SSH-like unit-cell}\label{sec:theo/mdl} 


%

\begin{figure}[ht]
	\includegraphics[width=0.8\linewidth]{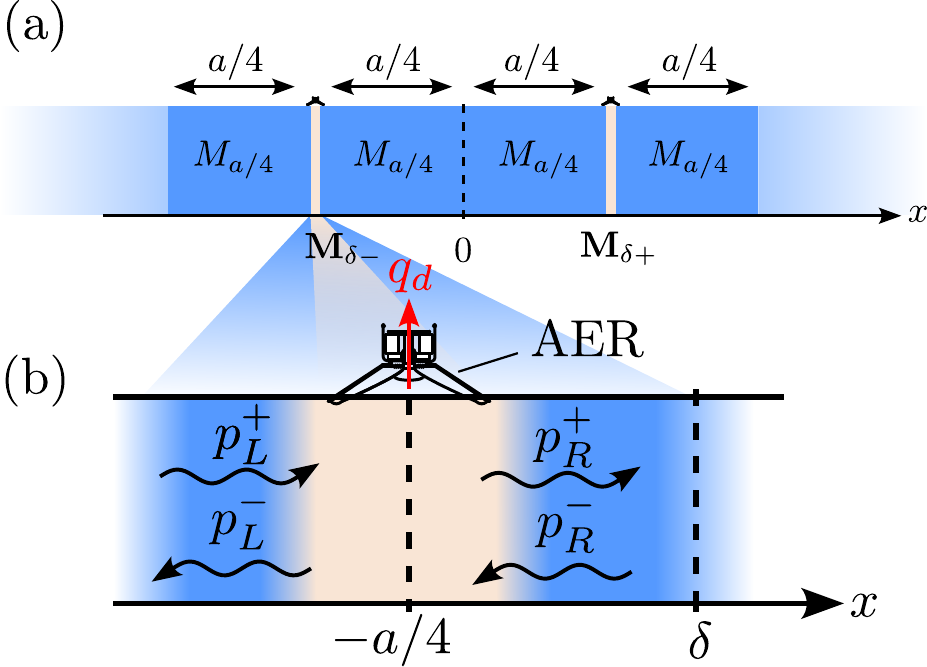}
	\caption{\label{fig:unitcell_theory} Transfer Matrix approach to the active acoustic SSH-like unit cell. (a) Unit cell of length $a$ composed of a duct segment loaded with two equally spaced AERs that are fixed in place. It can be split into 6 segments - each described by its own transfer matrix $M$. (b) An AER fixed at position $x = -a/4$. $\delta$ (- the dimerisation parameter) is the distance from the AER which is related to the coupling strength - c.f. main text.} Incoming and outgoing complex field amplitudes on both sides of the speaker plane interface $p_{R,L}^\pm$ are represented by the arrows.%
\end{figure}

Suppose a crystal unit cell composed of a duct segment loaded with two identical and equally spaced AERs that are fixed in place as shown in Figure~\ref{fig:unitcell_theory}a. A periodic array of these cells paired with coupling would form a SSH-like chain along the $x$ axis. Assuming continuity of pressure and volumetric flow, the transfer matrix $M_{cell}$ relating amplitudes in subsequent cells can be obtained using the Transfer Matrix Method (TMM) \cite{Richoux2002}. The latter solely depends on the cell's geometric properties and the complex mechanical impedances $\zeta_m$ of the resonators. The crystal band structure can be obtained by applying the Bloch theorem and computing the eigenvalues.

The total single cell transfer matrix of the active system is simply given by multiplying the individual transfer matrices of each of its constituents:
\begin{equation}\label{eq:Mcell}
	M_{cell} = M_{a/4}\mathbf{M_{\delta-}}M_{a/4}M_{a/4}\mathbf{M_{\delta+}}M_{a/4}	
\end{equation}
where $M_{a/4}$ represents the straight duct of length $L = a/4$ and $\mathbf{M_{\delta\pm}}$ is the transfer matrix through the AER plane with dimerisation parameter $\delta$ related to the cell coupling strength. 
Take the left AER which is placed flat against the duct wall at position $x=-a/4$ as shown in Figure~\ref{fig:unitcell_theory}b. $\mathbf{M_{\delta-}}$ is computed considering the following:
\begin{itemize}
	\item 
		The target \textit{remote} impedance is defined as:  
		\begin{equation}\label{eq:virt_impedance}
			\tilde{\zeta}_{\delta-} = S_d^2 \cdot \frac{p(x = -a/4 + \delta)}{q_d} 
		\end{equation}	
		where $S_d$, $p$, $q_d$ are the effective diaphragm surface area, pressure and volumetric flow through the diaphragm respectively.  In contrast to the \textit{real} speaker impedance, the evaluated pressure and velocity are here spatially separated by a distance $\delta$.\\
	\item		
		The pressure radiated from the speaker at any position $x$ in the duct can be written as: 
		\begin{equation}\label{eq:radiated_pressure}
			p_{rs}(x) = -\frac{\rho c q_d}{2S}e^{-ik|x+a/4|}
		\end{equation}
		where $\rho$, $c$, $S$, $k$ are the air density, speed of sound in the air, duct cross-sectional area and the wave number respectively. \\
	\item			
		The volumetric flow conservation and pressure continuity at $x=-a/4$ yields:
		\begin{align}\label{eq:flow_cons1}
			p_R^+ & = -\frac{\rho c}{2S}q_d + p_L^+\\ 
			p_L^- & = -\frac{\rho c}{2S}q_d + p_R^- \label{eq:flow_cons2}
		\end{align}
		where $p_L^+$, $p_L^-$ are the forward and backward complex field amplitudes on the left of the speaker and $p_R^+$, $p_R^-$ are the amplitudes on the right.	
\end{itemize}
%

The impedance defined in Eq.\eqref{eq:virt_impedance} is imposed by means of active control at the AER. Using equations \eqref{eq:radiated_pressure}, \eqref{eq:flow_cons1} and \eqref{eq:flow_cons2}, the transfer matrix for the left AER $\mathbf{M_{\delta-}}$ is obtained: 
\begin{equation*}
	\left(
	\begin{array}{l}
	p_{R}^{+} \\
	p_{R}^{-}
	\end{array}
	\right)=\underbrace{\left(\mathds{1}+ w_{\delta-}\cdot\left(
	\begin{array}{cc}
		e^{-jk\delta} & e^{+jk\delta} \\
		-e^{-jk\delta} & e^{+jk\delta}
	\end{array}
	\right)\right)^{-1}}_{\textstyle \mathbf{M_{\delta-}}}\left(
	\begin{array}{l}
		p_{L}^{+} \\
		p_{L}^{-}
	\end{array}\right),
\end{equation*}
\begin{equation}
	w_{\delta-}  = \frac{\rho c S_d^2}{2S \tilde{\zeta}_{\delta-}}
\end{equation}
By symmetry of the unit cell, the transfer matrix for the right AER, $M_{\delta+}$, is the same albeit with $\delta$ of opposite sign. 

Finally, the band structure is obtained using Eq.~\refeq{eq:Mcell} and applying the Floquet-Bloch theorem yields:

\begin{equation}
	q_F(k)= \arccos\left(\frac{1}{2} Tr(M_{cell}(k)) \right)/a
\end{equation}
where $q_F$ is the Floquet-Bloch wave number (- details in Appendix~\ref{app:disp}).

\section{Tailoring the dispersion through active control}
\begin{figure}[ht]
	\includegraphics[width =0.45\textwidth]{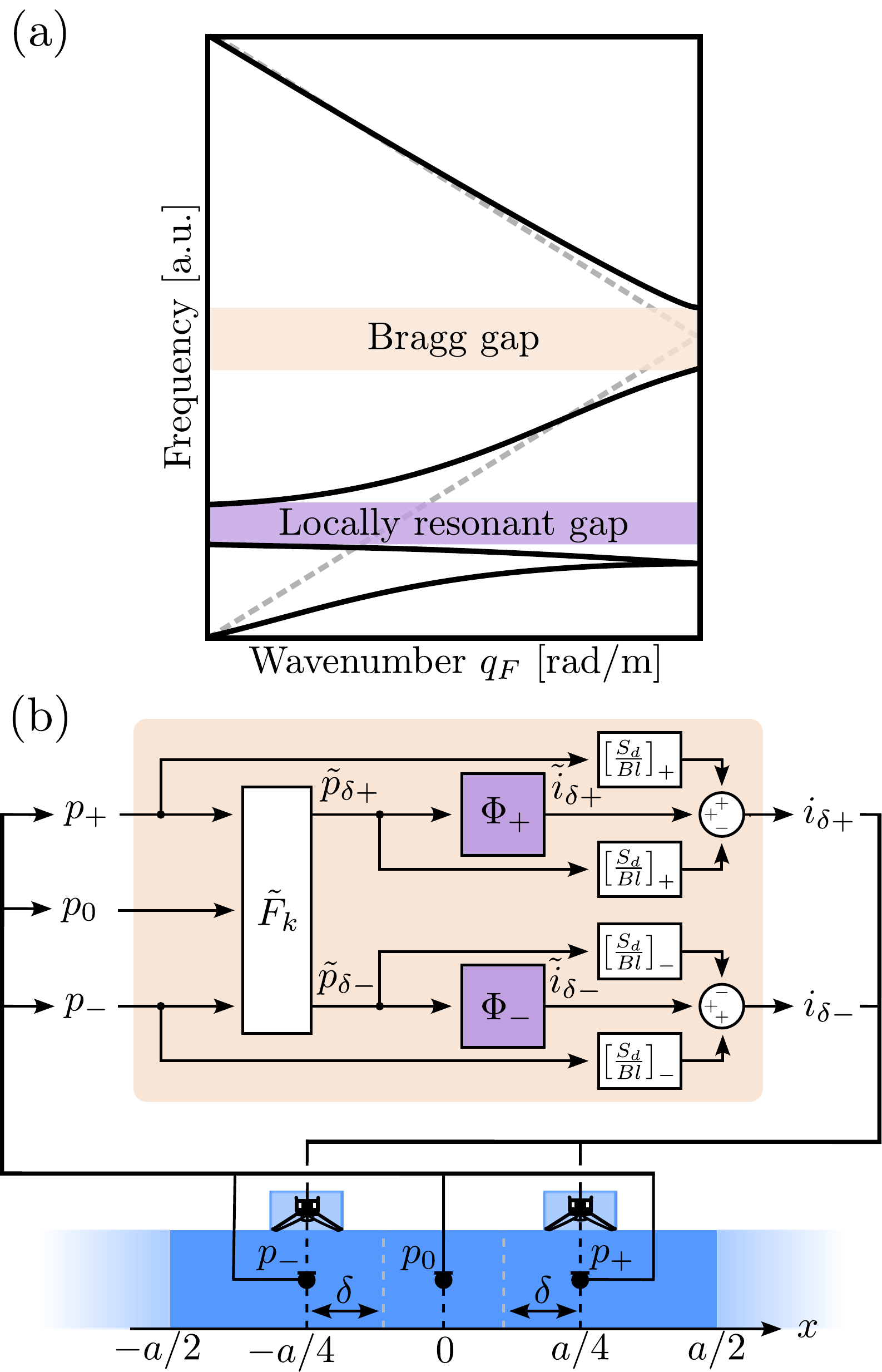}
	\caption{Active dispersion shaping. (a) Dispersion of a lossless coupled resonator system with two gaps: a Bragg gap and a locally resonant gap. (b) Block diagram of the proposed control scheme wired to the experimental unit cell. From left to right, the acoustic pressures in front of and in between the AERs (- $p_\pm$ and $p_0$ respectively) are used as inputs of the $\tilde{F}_k$ function which serves to achieve the $\delta$ shift. The impedance is synthesized with the desired mechanical parameters specified by the transfer functions $\Phi_\pm$. The output currents $i_{\delta\pm}$ are carried out to the respective AERs in the unit cell. Black dots in the unit cell are pressure microphones. }
	\label{fig:full_setup}
\end{figure}

In order to achieve active impedance control, the experimental unit cell is equipped with microphones placed in front and in between of the lined AERs as shown at the bottom of Fig.~\ref{fig:full_setup}b. The system is tuned via an active feedback control scheme which is split in two parts - each of which aims at altering the gaps present in the dispersion of the coupled resonator meta-structure, namely a band folding {\it Bragg gap} and a {\it locally resonant gap}, highlighted on Fig.~\ref{fig:full_setup}a.

\subsubsection{Shaping the locally resonant gap} 
The band-folding \textit{locally resonant gap} results from the coupling of a continuum of propagating waves and a local resonance which can be described by a polariton-like dispersion \cite{Yves2017} - albeit folded for a doubled unit cell.  Its position and size can be shaped at the experimenter's discretion by altering the AER's resonance frequency and quality factor \cite{Rivet2017, Lissek2018, Guo2020}. The desired mechanical impedance $\zeta_{st}$ is specified by control parameters $\mu_M,\mu_R,\mu_C$: 
 \begin{equation}\label{eq:mech_impedance}
 	\zeta_{st}(s) = \mu_M M_{ms}\cdot s+\mu_R R_{ms} + \mu_C/(C_{mc}\cdot s)
 \end{equation}
 
 Where:
 \begin{itemize}
 	\item $s = j\omega$: Laplace variable
 	\item $M_{ms}$: diaphragm mass (kg)
 	\item $R_{ms}$: mechanical resistance  (kg.s$^{-1}$)
 	\item $C_{mc}$: diaphragm + cabinet compliance (s$^2$.kg$^{-1}$)
 \end{itemize}

$\zeta_{st}$ synthesis is achieved by multiplying a transfer function $\Phi(\zeta_{st})$ \cite{Rivet2017a} to the real-time front pressure resulting in the current required to drive the AERs (- details in Appendix~\ref{app:impedance_synth}). 

 
\subsubsection{Shaping the Bragg gap} 

The band-folding \textit{Bragg gap} opens when dimerisation is introduced in the crystal \cite{Su1979}.  Here we show that a gap forms when the coupling is enhanced via active control. Figure~\ref{fig:full_setup}b shows the proposed block diagram of the control scheme. Knowing the operating wavenumber $k$ in the waveguide, the pressure $\tilde{p}_{\delta\pm}$ between an AER and the centre microphones can be estimated using the function:
\begin{equation}
	\tilde{F}_k(p_\pm,p_0) \coloneqq \frac{ p_\pm \cdot\sin(\pm k(a/4-\delta)) + p_0 \cdot\sin(\pm k\delta)}{\sin(\pm ka/4)}
\end{equation}
where $\delta$ is the dimerisation parameter.

 The virtual control current required to synthesize a target impedance $Z_{st}$ is:
\begin{equation}
	\tilde{i}_{\delta\pm} = 	\tilde{p}_{\delta\pm}\cdot\Phi_\pm
\end{equation}
where the $\Phi_\pm$ is the aforementioned transfer function. Notice that $\Phi$ appears twice in the control scheme block diagram as the AERs mechanical parameters are synthesized separately.

Finally, the control current $i_{\delta\pm}$ used to drive the AERs are:
\begin{equation}
	i_{\delta\pm} = (p_\pm - \tilde{p}_{\delta\pm})\cdot \left[\frac{S_d}{Bl}\right]_\pm + \tilde{i}_{\delta\pm}
\end{equation}

\section{Results and Discussion}\label{sec:obj/results}

\subsection{Analytical model}\label{sec:obj/mdl}

 The theoretical dispersion is computed using Eq.~\eqref{eq:Mcell}:
\begin{widetext}\label{eq::q_F}
	\begin{equation} 
		\cos (q_F (k)a) = \frac{w_{\delta+} w_{\delta-}[\cos (ka) - \cos(k 2\delta )] + w_{\delta+} j \sin (k(a+\delta))+ w_{\delta-}   j \sin (k(a-\delta)) + 	\cos(ka)}{j(w_{\delta+} 2 j \sin(k \delta)+1)(w_{\delta-}  2 j \sin(k \delta)-1)}
	\end{equation}
\end{widetext}

\begin{figure}[!h]
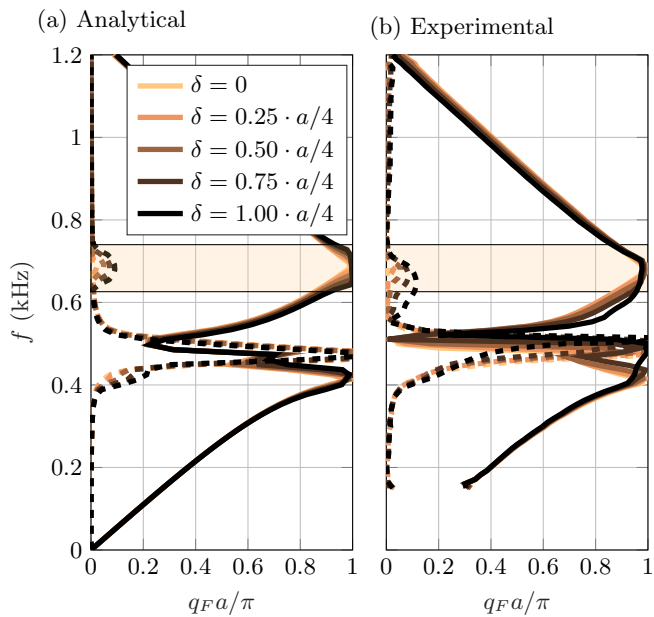

	\begin{minipage}[b]{0.21\textwidth}
 	\begin{flushleft}
	\hspace{10pt} (a) Analytical
		\end{flushleft}
		\vspace{-3mm} 
		\input{./figures/tikz/disp_ana.tex}
	\end{minipage}
	\hfill
	\begin{minipage}[b]{0.21\textwidth}
  \begin{flushleft}
			(b) Experimental
		\end{flushleft}
        \vspace{-3mm} 
		\input{./figures/tikz/disp_exp.tex}
	\end{minipage}
	\caption{Active control on the band gap size. Top: The dispersion diagrams of the acoustic SSH-like for five values of the dimerisation parameter $\delta$ derived from the analytical model (a) and measured experimental data (b). Solid lines and dashed lines represent the real and imaginary parts of the Bloch-Floquet wave number $q_F$ respectively. The largest Bragg gap is indicated by the orange transparent zones.}\label{fig:disp}
\end{figure}

The latter is plotted for dimerisation values $\delta \in (0,+a/4)$ as shown in Figure~\ref{fig:disp}a. The locally resonant and Bragg gaps occur at 500 Hz and 690 Hz respectively. For $\delta = 0$, the Bragg gap is closed as indicated by the vanishing complex part of the Floquet-Bloch wavenumber $q_F$. This comes as no surprise since this corresponds to the dispersion of equally spaced coupled resonators \cite{Zhao2021}. As $\delta$ increases however, the gap opens as a result of the change of the \textit{intra}-cell coupling. Conversely, negative values of $\delta$ corresponding to \textit{inter}-cell coupling yield identical dispersions.

\subsection{Experimental demonstration}\label{sec:obj/exp}

\begin{figure}[ht]
	\includegraphics[width=0.8\linewidth]{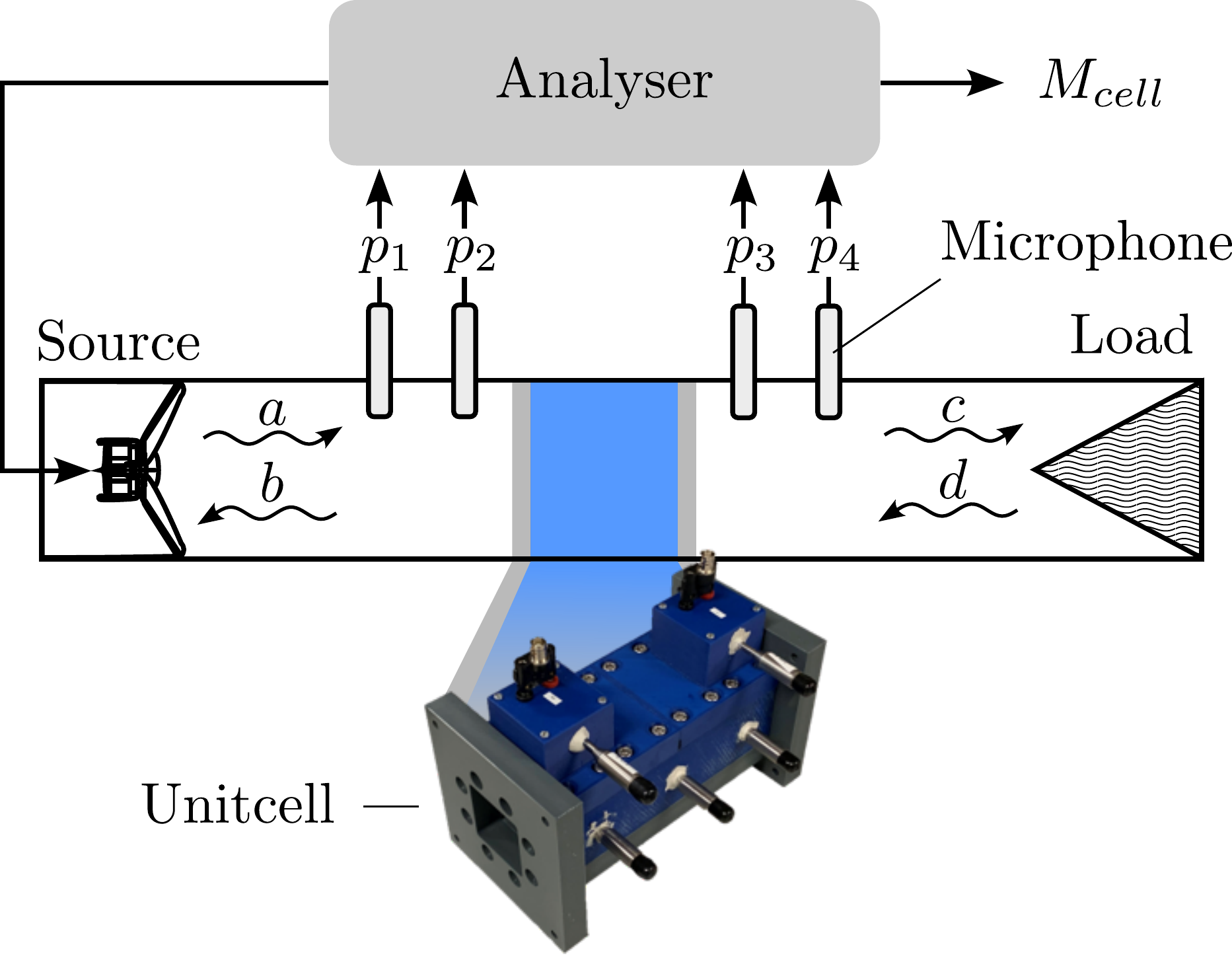}
	\caption{Characterisation of the experimental unit cell using a loaded Kundt duct and four pressure microphones. Spatial separation of the microphones on both sides of the sample allows for the determination of the forward ($a$,$c$) and backward ($b$,$d$) complex field amplitudes and subsequently the sample transfer matrix $M_{cell}$ via a digital analyser.}%
	\label{fig:exp_setup}
\end{figure}

One active SSH-like unit-cell was successfully designed, 3D-printed and assembled. The experimental AERs are off-the-shelf Vistaton FRWS 5 8 $\Omega$ drivers. Characterization of the mechanical parameters of the latter was achieved by direct measurement of the impedance following methods described by E. Rivet \cite{Rivet2017a} using quarter-inch PCB microphones and a single-point Polytec laser vibrometer for simultaneous pressure and velocity measurement respectively. Active control is carried out by  commercial Performance Speedgoat machine equipped with the IO334 module (- see Appendix~\ref{app:impedance_synth}). COMSOL Multiphysics finite element simulations were performed enabling design optimization and experimental validation. The mass, resistance and compliance were characterised and actively tuned as summarised in the Table~\ref{tab:imp_synth}. The goal was first to decrease the resonance frequency and second, to increase the quality factor of the resonators. The former effectively increases the locally resonant gap and Bragg gap frequency separation and the latter decreases system losses.  
\begingroup
\setlength{\tabcolsep}{10pt} 
\renewcommand{\arraystretch}{1.5} 
\begin{table}[ht]
		\begin{tabular}{ c c c c }
			& Mass & Resistance & Compliance \\ \hline
			Nominal       & 5.29 g & 0.297 kg/s      & 175 $\mu$m/N      \\
			$\mu_{M,R,C}$ & 0.5    & 0.09      & 0.42      		\\ \hline
			Synthesized   & 2.65 g & 0.0267 kg/s    & 73.5 $\mu$m/N     
		\end{tabular}%
\caption{\label{tab:imp_synth} Impedance synthesis. The nominal/passive mechanical parameters of the AER are multiplided by the respective tuneing parameters $\mu_{M,R,C}$ to obtain the desired synthesised mechanical parameters.}
\end{table}
\endgroup

 Experimental dispersion curves were obtained using the aforementioned TMM. The total unit cell transfer matrix $M_{cell}$ is measured following the ASTM standard procedure \cite{Standards2011} which relies on impedance tube measurements with four microphones as shown in Figure~\ref{fig:exp_setup}. The experimental dispersion curves are plotted in Figure~\ref{fig:disp}b.

Remarkable consistency can be noticed throughout both the model and the experimental results reflecting the efficacy of the proposed active control scheme. In particular, the Bragg gaps open as a function of the dimerisation parameter $\delta$. There are however some noticeable discrepancies around the natural resonance frequency of the resonators. Indeed, reliable feed-forward impedance synthesis demands highly accurate estimations of the ERs mechanical parameters - this is especially relevant for the heavy synthesis used here. Misestimations of these parameters can result in synthesizing multiple-degrees-of-freedom impedances which is perhaps made evident by an increased evanescent component $\Im (q_F)$ of the wave around the local resonance in the experimental dispersion. More accurate impedance synthesis control schemes may be worth investigating \cite{Guo2022,Volery2023} to mitigate these effects. That being said, this high energy dissipation at the resonance has little, if any, effect on the Bragg gap which is the focus of this work.\\







\subsection{Topological interface states in a finite  meta-crystal}\label{sec:obj/topo}
The topological nature of the system can be made evident by interfacing two topologically distinct phases of the crystal and seeking out interface states. The latter will appear as a transmission peak within the aforementioned Bragg gap.  An 8-cell system has been modelled using COMSOL Multiphysics (- Fig.~\ref{fig::8cell_sim_all}a) for multiple coupling configurations. The finite crystal is composed of two parts labelled $A$ and $B$ where the dimerisation parameters, $\delta_A$ and $\delta_B$, are independently controlled. By setting ports at the extremities of the crystal, the scattering matrix $S$ and consequently the transmission $T$ and reflection $R$ can be computed. Figure~\ref{fig::8cell_sim_all}b shows $R$ and $T$ as function of frequency for the following coupling configurations:

\begin{figure}[ht]
	\begin{minipage}[b]{0.5\textwidth}
		\begin{flushleft}
			(a)
		\end{flushleft}
		\vspace{-5mm} 
		\includegraphics[width =\textwidth]{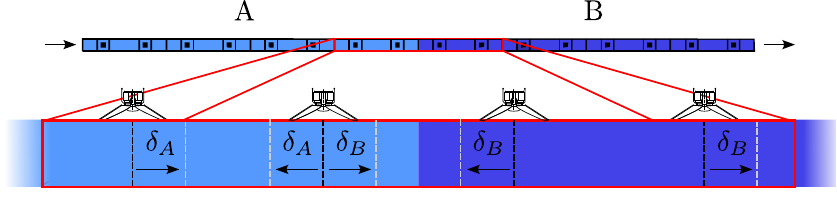}
	\end{minipage}\\
\bigskip
	\begin{minipage}[b]{0.53\textwidth}
		\begin{flushleft}
			(b)
		\end{flushleft}
		\vspace{-8mm} 
		\input{./figures/tikz/8cell_sim_all.tex}
		
	\end{minipage}
	\caption{\label{fig::8cell_sim_all} Simulated topological interface states in an 8-active-dimer system. (a) Finite element model of the 8-cell system with active dimers. The finite crystal is composed of two parts labelled $A$ and $B$ where the dimerisation parameters are independently controlled. (b) Transmission $T$ and reflection $R$ as a function of frequency for different coupling configurations. The dispersion Bragg gap corresponding to $\delta = \pm a/4$ from the analytical model is indicated by the orange transparent bars.}
\end{figure}

\begin{figure}[ht]
	\begin{minipage}[b]{0.48\textwidth}
		\begin{flushleft}
			(a)
		\end{flushleft}
		\vspace{-5mm} 
		\includegraphics[width =\textwidth]{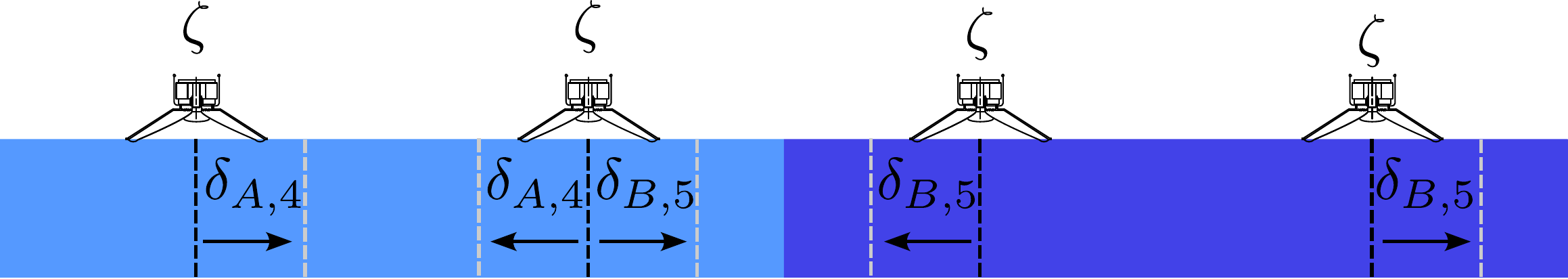}
	\end{minipage}\\
	\begin{minipage}[b]{0.45\textwidth}
%
%
\begin{tikzpicture}

\begin{axis}[%
width= 0.45\textwidth,
height=0.35\textwidth,
at={(0\textwidth,0pt)},
scale only axis,
point meta min=0.151067230066183,
point meta max=0.73371719026832,
axis on top,
xmin=0.65,
xmax=0.75,
xlabel style={font=\color{white!15!black}},
xlabel={f (kHz)},
ymin=0,
ymax=40,
ylabel style={font=\color{white!15!black}},
ylabel={DS/a (\%)},
axis background/.style={fill=white},
axis x line*=bottom,
axis y line*=left,
xtick distance=0.05,
colormap={mymap}{[1pt] rgb(0pt)=(0.2422,0.1504,0.6603); rgb(1pt)=(0.2444,0.1534,0.6728); rgb(2pt)=(0.2464,0.1569,0.6847); rgb(3pt)=(0.2484,0.1607,0.6961); rgb(4pt)=(0.2503,0.1648,0.7071); rgb(5pt)=(0.2522,0.1689,0.7179); rgb(6pt)=(0.254,0.1732,0.7286); rgb(7pt)=(0.2558,0.1773,0.7393); rgb(8pt)=(0.2576,0.1814,0.7501); rgb(9pt)=(0.2594,0.1854,0.761); rgb(11pt)=(0.2628,0.1932,0.7828); rgb(12pt)=(0.2645,0.1972,0.7937); rgb(13pt)=(0.2661,0.2011,0.8043); rgb(14pt)=(0.2676,0.2052,0.8148); rgb(15pt)=(0.2691,0.2094,0.8249); rgb(16pt)=(0.2704,0.2138,0.8346); rgb(17pt)=(0.2717,0.2184,0.8439); rgb(18pt)=(0.2729,0.2231,0.8528); rgb(19pt)=(0.274,0.228,0.8612); rgb(20pt)=(0.2749,0.233,0.8692); rgb(21pt)=(0.2758,0.2382,0.8767); rgb(22pt)=(0.2766,0.2435,0.884); rgb(23pt)=(0.2774,0.2489,0.8908); rgb(24pt)=(0.2781,0.2543,0.8973); rgb(25pt)=(0.2788,0.2598,0.9035); rgb(26pt)=(0.2794,0.2653,0.9094); rgb(27pt)=(0.2798,0.2708,0.915); rgb(28pt)=(0.2802,0.2764,0.9204); rgb(29pt)=(0.2806,0.2819,0.9255); rgb(30pt)=(0.2809,0.2875,0.9305); rgb(31pt)=(0.2811,0.293,0.9352); rgb(32pt)=(0.2813,0.2985,0.9397); rgb(33pt)=(0.2814,0.304,0.9441); rgb(34pt)=(0.2814,0.3095,0.9483); rgb(35pt)=(0.2813,0.315,0.9524); rgb(36pt)=(0.2811,0.3204,0.9563); rgb(37pt)=(0.2809,0.3259,0.96); rgb(38pt)=(0.2807,0.3313,0.9636); rgb(39pt)=(0.2803,0.3367,0.967); rgb(40pt)=(0.2798,0.3421,0.9702); rgb(41pt)=(0.2791,0.3475,0.9733); rgb(42pt)=(0.2784,0.3529,0.9763); rgb(43pt)=(0.2776,0.3583,0.9791); rgb(44pt)=(0.2766,0.3638,0.9817); rgb(45pt)=(0.2754,0.3693,0.984); rgb(46pt)=(0.2741,0.3748,0.9862); rgb(47pt)=(0.2726,0.3804,0.9881); rgb(48pt)=(0.271,0.386,0.9898); rgb(49pt)=(0.2691,0.3916,0.9912); rgb(50pt)=(0.267,0.3973,0.9924); rgb(51pt)=(0.2647,0.403,0.9935); rgb(52pt)=(0.2621,0.4088,0.9946); rgb(53pt)=(0.2591,0.4145,0.9955); rgb(54pt)=(0.2556,0.4203,0.9965); rgb(55pt)=(0.2517,0.4261,0.9974); rgb(56pt)=(0.2473,0.4319,0.9983); rgb(57pt)=(0.2424,0.4378,0.9991); rgb(58pt)=(0.2369,0.4437,0.9996); rgb(59pt)=(0.2311,0.4497,0.9995); rgb(60pt)=(0.225,0.4559,0.9985); rgb(61pt)=(0.2189,0.462,0.9968); rgb(62pt)=(0.2128,0.4682,0.9948); rgb(63pt)=(0.2066,0.4743,0.9926); rgb(64pt)=(0.2006,0.4803,0.9906); rgb(65pt)=(0.195,0.4861,0.9887); rgb(66pt)=(0.1903,0.4919,0.9867); rgb(67pt)=(0.1869,0.4975,0.9844); rgb(68pt)=(0.1847,0.503,0.9819); rgb(69pt)=(0.1831,0.5084,0.9793); rgb(70pt)=(0.1818,0.5138,0.9766); rgb(71pt)=(0.1806,0.5191,0.9738); rgb(72pt)=(0.1795,0.5244,0.9709); rgb(73pt)=(0.1785,0.5296,0.9677); rgb(74pt)=(0.1778,0.5349,0.9641); rgb(75pt)=(0.1773,0.5401,0.9602); rgb(76pt)=(0.1768,0.5452,0.956); rgb(77pt)=(0.1764,0.5504,0.9516); rgb(78pt)=(0.1755,0.5554,0.9473); rgb(79pt)=(0.174,0.5605,0.9432); rgb(80pt)=(0.1716,0.5655,0.9393); rgb(81pt)=(0.1686,0.5705,0.9357); rgb(82pt)=(0.1649,0.5755,0.9323); rgb(83pt)=(0.161,0.5805,0.9289); rgb(84pt)=(0.1573,0.5854,0.9254); rgb(85pt)=(0.154,0.5902,0.9218); rgb(86pt)=(0.1513,0.595,0.9182); rgb(87pt)=(0.1492,0.5997,0.9147); rgb(88pt)=(0.1475,0.6043,0.9113); rgb(89pt)=(0.1461,0.6089,0.908); rgb(90pt)=(0.1446,0.6135,0.905); rgb(91pt)=(0.1429,0.618,0.9022); rgb(92pt)=(0.1408,0.6226,0.8998); rgb(93pt)=(0.1383,0.6272,0.8975); rgb(94pt)=(0.1354,0.6317,0.8953); rgb(95pt)=(0.1321,0.6363,0.8932); rgb(96pt)=(0.1288,0.6408,0.891); rgb(97pt)=(0.1253,0.6453,0.8887); rgb(98pt)=(0.1219,0.6497,0.8862); rgb(99pt)=(0.1185,0.6541,0.8834); rgb(100pt)=(0.1152,0.6584,0.8804); rgb(101pt)=(0.1119,0.6627,0.877); rgb(102pt)=(0.1085,0.6669,0.8734); rgb(103pt)=(0.1048,0.671,0.8695); rgb(104pt)=(0.1009,0.675,0.8653); rgb(105pt)=(0.0964,0.6789,0.8609); rgb(106pt)=(0.0914,0.6828,0.8562); rgb(107pt)=(0.0855,0.6865,0.8513); rgb(108pt)=(0.0789,0.6902,0.8462); rgb(109pt)=(0.0713,0.6938,0.8409); rgb(110pt)=(0.0628,0.6972,0.8355); rgb(111pt)=(0.0535,0.7006,0.8299); rgb(112pt)=(0.0433,0.7039,0.8242); rgb(113pt)=(0.0328,0.7071,0.8183); rgb(114pt)=(0.0234,0.7103,0.8124); rgb(115pt)=(0.0155,0.7133,0.8064); rgb(116pt)=(0.0091,0.7163,0.8003); rgb(117pt)=(0.0046,0.7192,0.7941); rgb(118pt)=(0.0019,0.722,0.7878); rgb(119pt)=(0.0009,0.7248,0.7815); rgb(120pt)=(0.0018,0.7275,0.7752); rgb(121pt)=(0.0046,0.7301,0.7688); rgb(122pt)=(0.0094,0.7327,0.7623); rgb(123pt)=(0.0162,0.7352,0.7558); rgb(124pt)=(0.0253,0.7376,0.7492); rgb(125pt)=(0.0369,0.74,0.7426); rgb(126pt)=(0.0504,0.7423,0.7359); rgb(127pt)=(0.0638,0.7446,0.7292); rgb(128pt)=(0.077,0.7468,0.7224); rgb(129pt)=(0.0899,0.7489,0.7156); rgb(130pt)=(0.1023,0.751,0.7088); rgb(131pt)=(0.1141,0.7531,0.7019); rgb(132pt)=(0.1252,0.7552,0.695); rgb(133pt)=(0.1354,0.7572,0.6881); rgb(134pt)=(0.1448,0.7593,0.6812); rgb(135pt)=(0.1532,0.7614,0.6741); rgb(136pt)=(0.1609,0.7635,0.6671); rgb(137pt)=(0.1678,0.7656,0.6599); rgb(138pt)=(0.1741,0.7678,0.6527); rgb(139pt)=(0.1799,0.7699,0.6454); rgb(140pt)=(0.1853,0.7721,0.6379); rgb(141pt)=(0.1905,0.7743,0.6303); rgb(142pt)=(0.1954,0.7765,0.6225); rgb(143pt)=(0.2003,0.7787,0.6146); rgb(144pt)=(0.2061,0.7808,0.6065); rgb(145pt)=(0.2118,0.7828,0.5983); rgb(146pt)=(0.2178,0.7849,0.5899); rgb(147pt)=(0.2244,0.7869,0.5813); rgb(148pt)=(0.2318,0.7887,0.5725); rgb(149pt)=(0.2401,0.7905,0.5636); rgb(150pt)=(0.2491,0.7922,0.5546); rgb(151pt)=(0.2589,0.7937,0.5454); rgb(152pt)=(0.2695,0.7951,0.536); rgb(153pt)=(0.2809,0.7964,0.5266); rgb(154pt)=(0.2929,0.7975,0.517); rgb(155pt)=(0.3052,0.7985,0.5074); rgb(156pt)=(0.3176,0.7994,0.4975); rgb(157pt)=(0.3301,0.8002,0.4876); rgb(158pt)=(0.3424,0.8009,0.4774); rgb(159pt)=(0.3548,0.8016,0.4669); rgb(160pt)=(0.3671,0.8021,0.4563); rgb(161pt)=(0.3795,0.8026,0.4454); rgb(162pt)=(0.3921,0.8029,0.4344); rgb(163pt)=(0.405,0.8031,0.4233); rgb(164pt)=(0.4184,0.803,0.4122); rgb(165pt)=(0.4322,0.8028,0.4013); rgb(166pt)=(0.4463,0.8024,0.3904); rgb(167pt)=(0.4608,0.8018,0.3797); rgb(168pt)=(0.4753,0.8011,0.3691); rgb(169pt)=(0.4899,0.8002,0.3586); rgb(170pt)=(0.5044,0.7993,0.348); rgb(171pt)=(0.5187,0.7982,0.3374); rgb(172pt)=(0.5329,0.797,0.3267); rgb(173pt)=(0.547,0.7957,0.3159); rgb(175pt)=(0.5748,0.7929,0.2941); rgb(176pt)=(0.5886,0.7913,0.2833); rgb(177pt)=(0.6024,0.7896,0.2726); rgb(178pt)=(0.6161,0.7878,0.2622); rgb(179pt)=(0.6297,0.7859,0.2521); rgb(180pt)=(0.6433,0.7839,0.2423); rgb(181pt)=(0.6567,0.7818,0.2329); rgb(182pt)=(0.6701,0.7796,0.2239); rgb(183pt)=(0.6833,0.7773,0.2155); rgb(184pt)=(0.6963,0.775,0.2075); rgb(185pt)=(0.7091,0.7727,0.1998); rgb(186pt)=(0.7218,0.7703,0.1924); rgb(187pt)=(0.7344,0.7679,0.1852); rgb(188pt)=(0.7468,0.7654,0.1782); rgb(189pt)=(0.759,0.7629,0.1717); rgb(190pt)=(0.771,0.7604,0.1658); rgb(191pt)=(0.7829,0.7579,0.1608); rgb(192pt)=(0.7945,0.7554,0.157); rgb(193pt)=(0.806,0.7529,0.1546); rgb(194pt)=(0.8172,0.7505,0.1535); rgb(195pt)=(0.8281,0.7481,0.1536); rgb(196pt)=(0.8389,0.7457,0.1546); rgb(197pt)=(0.8495,0.7435,0.1564); rgb(198pt)=(0.86,0.7413,0.1587); rgb(199pt)=(0.8703,0.7392,0.1615); rgb(200pt)=(0.8804,0.7372,0.165); rgb(201pt)=(0.8903,0.7353,0.1695); rgb(202pt)=(0.9,0.7336,0.1749); rgb(203pt)=(0.9093,0.7321,0.1815); rgb(204pt)=(0.9184,0.7308,0.189); rgb(205pt)=(0.9272,0.7298,0.1973); rgb(206pt)=(0.9357,0.729,0.2061); rgb(207pt)=(0.944,0.7285,0.2151); rgb(208pt)=(0.9523,0.7284,0.2237); rgb(209pt)=(0.9606,0.7285,0.2312); rgb(210pt)=(0.9689,0.7292,0.2373); rgb(211pt)=(0.977,0.7304,0.2418); rgb(212pt)=(0.9842,0.733,0.2446); rgb(213pt)=(0.99,0.7365,0.2429); rgb(214pt)=(0.9946,0.7407,0.2394); rgb(215pt)=(0.9966,0.7458,0.2351); rgb(216pt)=(0.9971,0.7513,0.2309); rgb(217pt)=(0.9972,0.7569,0.2267); rgb(218pt)=(0.9971,0.7626,0.2224); rgb(219pt)=(0.9969,0.7683,0.2181); rgb(220pt)=(0.9966,0.774,0.2138); rgb(221pt)=(0.9962,0.7798,0.2095); rgb(222pt)=(0.9957,0.7856,0.2053); rgb(223pt)=(0.9949,0.7915,0.2012); rgb(224pt)=(0.9938,0.7974,0.1974); rgb(225pt)=(0.9923,0.8034,0.1939); rgb(226pt)=(0.9906,0.8095,0.1906); rgb(227pt)=(0.9885,0.8156,0.1875); rgb(228pt)=(0.9861,0.8218,0.1846); rgb(229pt)=(0.9835,0.828,0.1817); rgb(230pt)=(0.9807,0.8342,0.1787); rgb(231pt)=(0.9778,0.8404,0.1757); rgb(232pt)=(0.9748,0.8467,0.1726); rgb(233pt)=(0.972,0.8529,0.1695); rgb(234pt)=(0.9694,0.8591,0.1665); rgb(235pt)=(0.9671,0.8654,0.1636); rgb(236pt)=(0.9651,0.8716,0.1608); rgb(237pt)=(0.9634,0.8778,0.1582); rgb(238pt)=(0.9619,0.884,0.1557); rgb(239pt)=(0.9608,0.8902,0.1532); rgb(240pt)=(0.9601,0.8963,0.1507); rgb(241pt)=(0.9596,0.9023,0.148); rgb(242pt)=(0.9595,0.9084,0.145); rgb(243pt)=(0.9597,0.9143,0.1418); rgb(244pt)=(0.9601,0.9203,0.1382); rgb(245pt)=(0.9608,0.9262,0.1344); rgb(246pt)=(0.9618,0.932,0.1304); rgb(247pt)=(0.9629,0.9379,0.1261); rgb(248pt)=(0.9642,0.9437,0.1216); rgb(249pt)=(0.9657,0.9494,0.1168); rgb(250pt)=(0.9674,0.9552,0.1116); rgb(251pt)=(0.9692,0.9609,0.1061); rgb(252pt)=(0.9711,0.9667,0.1001); rgb(253pt)=(0.973,0.9724,0.0938); rgb(254pt)=(0.9749,0.9782,0.0872); rgb(255pt)=(0.9769,0.9839,0.0805)},
colorbar,
colorbar style={ylabel style={font=\color{white!15!black}}, ylabel={Transmission}}
]
\addplot [forget plot] graphics [xmin=0.6495, xmax=0.7505, ymin=-0.5, ymax=40.5] {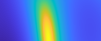};
\end{axis}

\end{tikzpicture}%
	\end{minipage}\\
\bigskip
	\begin{minipage}[b]{0.48\textwidth}
		\begin{flushleft}
			(b)
		\end{flushleft}
		\vspace{-5mm} 
	\includegraphics[width =\textwidth]{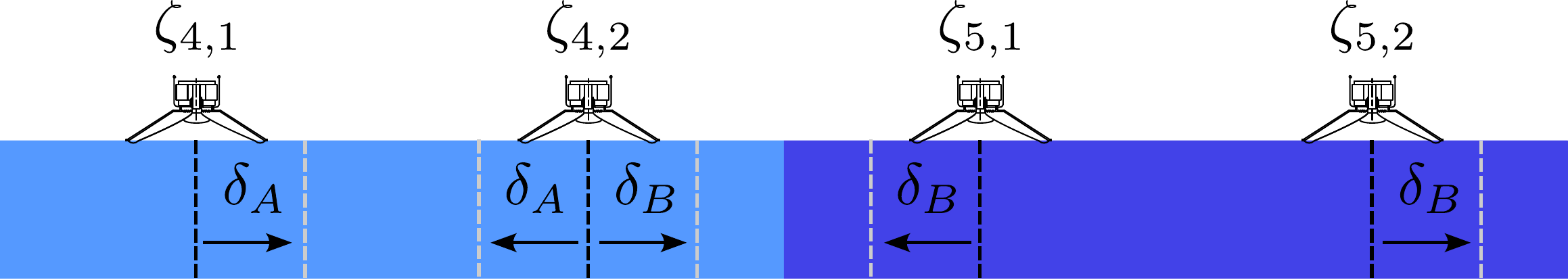}
\end{minipage}\\
	\begin{minipage}[b]{0.45\textwidth}
%
%
\begin{tikzpicture}

\begin{axis}[%
width= 0.45\textwidth,
height=0.35\textwidth,
at={(0\textwidth,0pt)},
scale only axis,
point meta min=0.112226528337832,
point meta max=0.733717190268317,
axis on top,
xmin=0.65,
xmax=0.75,
xlabel style={font=\color{white!15!black}},
xlabel={f (kHz)},
ymin=0,
ymax=40,
ylabel style={font=\color{white!15!black}},
ylabel={DS (\%)},
axis background/.style={fill=white},
axis x line*=bottom,
axis y line*=left,
xtick distance=0.05,
colormap={mymap}{[1pt] rgb(0pt)=(0.2422,0.1504,0.6603); rgb(1pt)=(0.2444,0.1534,0.6728); rgb(2pt)=(0.2464,0.1569,0.6847); rgb(3pt)=(0.2484,0.1607,0.6961); rgb(4pt)=(0.2503,0.1648,0.7071); rgb(5pt)=(0.2522,0.1689,0.7179); rgb(6pt)=(0.254,0.1732,0.7286); rgb(7pt)=(0.2558,0.1773,0.7393); rgb(8pt)=(0.2576,0.1814,0.7501); rgb(9pt)=(0.2594,0.1854,0.761); rgb(11pt)=(0.2628,0.1932,0.7828); rgb(12pt)=(0.2645,0.1972,0.7937); rgb(13pt)=(0.2661,0.2011,0.8043); rgb(14pt)=(0.2676,0.2052,0.8148); rgb(15pt)=(0.2691,0.2094,0.8249); rgb(16pt)=(0.2704,0.2138,0.8346); rgb(17pt)=(0.2717,0.2184,0.8439); rgb(18pt)=(0.2729,0.2231,0.8528); rgb(19pt)=(0.274,0.228,0.8612); rgb(20pt)=(0.2749,0.233,0.8692); rgb(21pt)=(0.2758,0.2382,0.8767); rgb(22pt)=(0.2766,0.2435,0.884); rgb(23pt)=(0.2774,0.2489,0.8908); rgb(24pt)=(0.2781,0.2543,0.8973); rgb(25pt)=(0.2788,0.2598,0.9035); rgb(26pt)=(0.2794,0.2653,0.9094); rgb(27pt)=(0.2798,0.2708,0.915); rgb(28pt)=(0.2802,0.2764,0.9204); rgb(29pt)=(0.2806,0.2819,0.9255); rgb(30pt)=(0.2809,0.2875,0.9305); rgb(31pt)=(0.2811,0.293,0.9352); rgb(32pt)=(0.2813,0.2985,0.9397); rgb(33pt)=(0.2814,0.304,0.9441); rgb(34pt)=(0.2814,0.3095,0.9483); rgb(35pt)=(0.2813,0.315,0.9524); rgb(36pt)=(0.2811,0.3204,0.9563); rgb(37pt)=(0.2809,0.3259,0.96); rgb(38pt)=(0.2807,0.3313,0.9636); rgb(39pt)=(0.2803,0.3367,0.967); rgb(40pt)=(0.2798,0.3421,0.9702); rgb(41pt)=(0.2791,0.3475,0.9733); rgb(42pt)=(0.2784,0.3529,0.9763); rgb(43pt)=(0.2776,0.3583,0.9791); rgb(44pt)=(0.2766,0.3638,0.9817); rgb(45pt)=(0.2754,0.3693,0.984); rgb(46pt)=(0.2741,0.3748,0.9862); rgb(47pt)=(0.2726,0.3804,0.9881); rgb(48pt)=(0.271,0.386,0.9898); rgb(49pt)=(0.2691,0.3916,0.9912); rgb(50pt)=(0.267,0.3973,0.9924); rgb(51pt)=(0.2647,0.403,0.9935); rgb(52pt)=(0.2621,0.4088,0.9946); rgb(53pt)=(0.2591,0.4145,0.9955); rgb(54pt)=(0.2556,0.4203,0.9965); rgb(55pt)=(0.2517,0.4261,0.9974); rgb(56pt)=(0.2473,0.4319,0.9983); rgb(57pt)=(0.2424,0.4378,0.9991); rgb(58pt)=(0.2369,0.4437,0.9996); rgb(59pt)=(0.2311,0.4497,0.9995); rgb(60pt)=(0.225,0.4559,0.9985); rgb(61pt)=(0.2189,0.462,0.9968); rgb(62pt)=(0.2128,0.4682,0.9948); rgb(63pt)=(0.2066,0.4743,0.9926); rgb(64pt)=(0.2006,0.4803,0.9906); rgb(65pt)=(0.195,0.4861,0.9887); rgb(66pt)=(0.1903,0.4919,0.9867); rgb(67pt)=(0.1869,0.4975,0.9844); rgb(68pt)=(0.1847,0.503,0.9819); rgb(69pt)=(0.1831,0.5084,0.9793); rgb(70pt)=(0.1818,0.5138,0.9766); rgb(71pt)=(0.1806,0.5191,0.9738); rgb(72pt)=(0.1795,0.5244,0.9709); rgb(73pt)=(0.1785,0.5296,0.9677); rgb(74pt)=(0.1778,0.5349,0.9641); rgb(75pt)=(0.1773,0.5401,0.9602); rgb(76pt)=(0.1768,0.5452,0.956); rgb(77pt)=(0.1764,0.5504,0.9516); rgb(78pt)=(0.1755,0.5554,0.9473); rgb(79pt)=(0.174,0.5605,0.9432); rgb(80pt)=(0.1716,0.5655,0.9393); rgb(81pt)=(0.1686,0.5705,0.9357); rgb(82pt)=(0.1649,0.5755,0.9323); rgb(83pt)=(0.161,0.5805,0.9289); rgb(84pt)=(0.1573,0.5854,0.9254); rgb(85pt)=(0.154,0.5902,0.9218); rgb(86pt)=(0.1513,0.595,0.9182); rgb(87pt)=(0.1492,0.5997,0.9147); rgb(88pt)=(0.1475,0.6043,0.9113); rgb(89pt)=(0.1461,0.6089,0.908); rgb(90pt)=(0.1446,0.6135,0.905); rgb(91pt)=(0.1429,0.618,0.9022); rgb(92pt)=(0.1408,0.6226,0.8998); rgb(93pt)=(0.1383,0.6272,0.8975); rgb(94pt)=(0.1354,0.6317,0.8953); rgb(95pt)=(0.1321,0.6363,0.8932); rgb(96pt)=(0.1288,0.6408,0.891); rgb(97pt)=(0.1253,0.6453,0.8887); rgb(98pt)=(0.1219,0.6497,0.8862); rgb(99pt)=(0.1185,0.6541,0.8834); rgb(100pt)=(0.1152,0.6584,0.8804); rgb(101pt)=(0.1119,0.6627,0.877); rgb(102pt)=(0.1085,0.6669,0.8734); rgb(103pt)=(0.1048,0.671,0.8695); rgb(104pt)=(0.1009,0.675,0.8653); rgb(105pt)=(0.0964,0.6789,0.8609); rgb(106pt)=(0.0914,0.6828,0.8562); rgb(107pt)=(0.0855,0.6865,0.8513); rgb(108pt)=(0.0789,0.6902,0.8462); rgb(109pt)=(0.0713,0.6938,0.8409); rgb(110pt)=(0.0628,0.6972,0.8355); rgb(111pt)=(0.0535,0.7006,0.8299); rgb(112pt)=(0.0433,0.7039,0.8242); rgb(113pt)=(0.0328,0.7071,0.8183); rgb(114pt)=(0.0234,0.7103,0.8124); rgb(115pt)=(0.0155,0.7133,0.8064); rgb(116pt)=(0.0091,0.7163,0.8003); rgb(117pt)=(0.0046,0.7192,0.7941); rgb(118pt)=(0.0019,0.722,0.7878); rgb(119pt)=(0.0009,0.7248,0.7815); rgb(120pt)=(0.0018,0.7275,0.7752); rgb(121pt)=(0.0046,0.7301,0.7688); rgb(122pt)=(0.0094,0.7327,0.7623); rgb(123pt)=(0.0162,0.7352,0.7558); rgb(124pt)=(0.0253,0.7376,0.7492); rgb(125pt)=(0.0369,0.74,0.7426); rgb(126pt)=(0.0504,0.7423,0.7359); rgb(127pt)=(0.0638,0.7446,0.7292); rgb(128pt)=(0.077,0.7468,0.7224); rgb(129pt)=(0.0899,0.7489,0.7156); rgb(130pt)=(0.1023,0.751,0.7088); rgb(131pt)=(0.1141,0.7531,0.7019); rgb(132pt)=(0.1252,0.7552,0.695); rgb(133pt)=(0.1354,0.7572,0.6881); rgb(134pt)=(0.1448,0.7593,0.6812); rgb(135pt)=(0.1532,0.7614,0.6741); rgb(136pt)=(0.1609,0.7635,0.6671); rgb(137pt)=(0.1678,0.7656,0.6599); rgb(138pt)=(0.1741,0.7678,0.6527); rgb(139pt)=(0.1799,0.7699,0.6454); rgb(140pt)=(0.1853,0.7721,0.6379); rgb(141pt)=(0.1905,0.7743,0.6303); rgb(142pt)=(0.1954,0.7765,0.6225); rgb(143pt)=(0.2003,0.7787,0.6146); rgb(144pt)=(0.2061,0.7808,0.6065); rgb(145pt)=(0.2118,0.7828,0.5983); rgb(146pt)=(0.2178,0.7849,0.5899); rgb(147pt)=(0.2244,0.7869,0.5813); rgb(148pt)=(0.2318,0.7887,0.5725); rgb(149pt)=(0.2401,0.7905,0.5636); rgb(150pt)=(0.2491,0.7922,0.5546); rgb(151pt)=(0.2589,0.7937,0.5454); rgb(152pt)=(0.2695,0.7951,0.536); rgb(153pt)=(0.2809,0.7964,0.5266); rgb(154pt)=(0.2929,0.7975,0.517); rgb(155pt)=(0.3052,0.7985,0.5074); rgb(156pt)=(0.3176,0.7994,0.4975); rgb(157pt)=(0.3301,0.8002,0.4876); rgb(158pt)=(0.3424,0.8009,0.4774); rgb(159pt)=(0.3548,0.8016,0.4669); rgb(160pt)=(0.3671,0.8021,0.4563); rgb(161pt)=(0.3795,0.8026,0.4454); rgb(162pt)=(0.3921,0.8029,0.4344); rgb(163pt)=(0.405,0.8031,0.4233); rgb(164pt)=(0.4184,0.803,0.4122); rgb(165pt)=(0.4322,0.8028,0.4013); rgb(166pt)=(0.4463,0.8024,0.3904); rgb(167pt)=(0.4608,0.8018,0.3797); rgb(168pt)=(0.4753,0.8011,0.3691); rgb(169pt)=(0.4899,0.8002,0.3586); rgb(170pt)=(0.5044,0.7993,0.348); rgb(171pt)=(0.5187,0.7982,0.3374); rgb(172pt)=(0.5329,0.797,0.3267); rgb(173pt)=(0.547,0.7957,0.3159); rgb(175pt)=(0.5748,0.7929,0.2941); rgb(176pt)=(0.5886,0.7913,0.2833); rgb(177pt)=(0.6024,0.7896,0.2726); rgb(178pt)=(0.6161,0.7878,0.2622); rgb(179pt)=(0.6297,0.7859,0.2521); rgb(180pt)=(0.6433,0.7839,0.2423); rgb(181pt)=(0.6567,0.7818,0.2329); rgb(182pt)=(0.6701,0.7796,0.2239); rgb(183pt)=(0.6833,0.7773,0.2155); rgb(184pt)=(0.6963,0.775,0.2075); rgb(185pt)=(0.7091,0.7727,0.1998); rgb(186pt)=(0.7218,0.7703,0.1924); rgb(187pt)=(0.7344,0.7679,0.1852); rgb(188pt)=(0.7468,0.7654,0.1782); rgb(189pt)=(0.759,0.7629,0.1717); rgb(190pt)=(0.771,0.7604,0.1658); rgb(191pt)=(0.7829,0.7579,0.1608); rgb(192pt)=(0.7945,0.7554,0.157); rgb(193pt)=(0.806,0.7529,0.1546); rgb(194pt)=(0.8172,0.7505,0.1535); rgb(195pt)=(0.8281,0.7481,0.1536); rgb(196pt)=(0.8389,0.7457,0.1546); rgb(197pt)=(0.8495,0.7435,0.1564); rgb(198pt)=(0.86,0.7413,0.1587); rgb(199pt)=(0.8703,0.7392,0.1615); rgb(200pt)=(0.8804,0.7372,0.165); rgb(201pt)=(0.8903,0.7353,0.1695); rgb(202pt)=(0.9,0.7336,0.1749); rgb(203pt)=(0.9093,0.7321,0.1815); rgb(204pt)=(0.9184,0.7308,0.189); rgb(205pt)=(0.9272,0.7298,0.1973); rgb(206pt)=(0.9357,0.729,0.2061); rgb(207pt)=(0.944,0.7285,0.2151); rgb(208pt)=(0.9523,0.7284,0.2237); rgb(209pt)=(0.9606,0.7285,0.2312); rgb(210pt)=(0.9689,0.7292,0.2373); rgb(211pt)=(0.977,0.7304,0.2418); rgb(212pt)=(0.9842,0.733,0.2446); rgb(213pt)=(0.99,0.7365,0.2429); rgb(214pt)=(0.9946,0.7407,0.2394); rgb(215pt)=(0.9966,0.7458,0.2351); rgb(216pt)=(0.9971,0.7513,0.2309); rgb(217pt)=(0.9972,0.7569,0.2267); rgb(218pt)=(0.9971,0.7626,0.2224); rgb(219pt)=(0.9969,0.7683,0.2181); rgb(220pt)=(0.9966,0.774,0.2138); rgb(221pt)=(0.9962,0.7798,0.2095); rgb(222pt)=(0.9957,0.7856,0.2053); rgb(223pt)=(0.9949,0.7915,0.2012); rgb(224pt)=(0.9938,0.7974,0.1974); rgb(225pt)=(0.9923,0.8034,0.1939); rgb(226pt)=(0.9906,0.8095,0.1906); rgb(227pt)=(0.9885,0.8156,0.1875); rgb(228pt)=(0.9861,0.8218,0.1846); rgb(229pt)=(0.9835,0.828,0.1817); rgb(230pt)=(0.9807,0.8342,0.1787); rgb(231pt)=(0.9778,0.8404,0.1757); rgb(232pt)=(0.9748,0.8467,0.1726); rgb(233pt)=(0.972,0.8529,0.1695); rgb(234pt)=(0.9694,0.8591,0.1665); rgb(235pt)=(0.9671,0.8654,0.1636); rgb(236pt)=(0.9651,0.8716,0.1608); rgb(237pt)=(0.9634,0.8778,0.1582); rgb(238pt)=(0.9619,0.884,0.1557); rgb(239pt)=(0.9608,0.8902,0.1532); rgb(240pt)=(0.9601,0.8963,0.1507); rgb(241pt)=(0.9596,0.9023,0.148); rgb(242pt)=(0.9595,0.9084,0.145); rgb(243pt)=(0.9597,0.9143,0.1418); rgb(244pt)=(0.9601,0.9203,0.1382); rgb(245pt)=(0.9608,0.9262,0.1344); rgb(246pt)=(0.9618,0.932,0.1304); rgb(247pt)=(0.9629,0.9379,0.1261); rgb(248pt)=(0.9642,0.9437,0.1216); rgb(249pt)=(0.9657,0.9494,0.1168); rgb(250pt)=(0.9674,0.9552,0.1116); rgb(251pt)=(0.9692,0.9609,0.1061); rgb(252pt)=(0.9711,0.9667,0.1001); rgb(253pt)=(0.973,0.9724,0.0938); rgb(254pt)=(0.9749,0.9782,0.0872); rgb(255pt)=(0.9769,0.9839,0.0805)},
colorbar,
colorbar style={ylabel style={font=\color{white!15!black}}, ylabel={Transmission}}
]
\addplot [forget plot] graphics [xmin=0.6495, xmax=0.7505, ymin=-0.5, ymax=40.5] {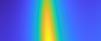};
\end{axis}

\end{tikzpicture}%
	\end{minipage}
	\caption{\label{fig:disorder}Effect of defects on the topological interface state. The centre transmission peak corresponding to the topological interface state is plotted in function the disorder strength (DS) for both coupling (a) and on-site (b) disorders. (a) All AERs along each part of the chain have identical mechanical impedances $\zeta$ and each cell coupling $\delta$ randomly (- uniformly) differs from one-another with strength DS. (b) The mechanical impedances $\zeta$ randomly differ and the cell couplings $\delta$ in each part are the same. }
\end{figure}


\begin{itemize}
\item $\delta_A =\delta_B = 0$: For equal coupling, the transmission is only impeded at the AER resonance frequency effectively drawing out the locally resonant gap.
 \item $\delta_A = \delta_B = +a/4$: The dimers of both parts are uniformly coupled resulting in an interface-less crystal of trivial phase. As expected from the dispersion curves, the transmission drops at the Bragg gap. Furthermore, the uniform topological phase ($\delta_A = \delta_B = - a/4$) yields similar results. 
 \item  $\delta_A = -\delta_B = +a/4$: An interface between trivial and topological phase is created. The latter yields a narrow transmission peak at the centre of the Bragg gap which is indicative of topological interface state.
\end{itemize}
 
The robustness of these topological states can be observed by tracing out the centre transmission peak when progressively generating coupling-parameter disorder in each cell. The latter is shown in Figure~\ref{fig:disorder}a where the the transmission peak remains almost unaffected by large disorders that do not close the band gap(- up to 25\% of lattice period).

\subsection{Symmetry protected topological interface states}\label{sec:obj/protected}
Generally speaking, one-dimensional topological phases are symmetry protected and here, one may be tempted to think the protection is a direct consequence of chiral symmetry as is the case in regular tight-binding SSH chains made of evanescently coupled identical resonators. However,  it is not the case for this acoustic system as the topological interface state is also robust against variations in AER resonance frequencies $\omega_0$ - i.e. on-site disorder. Figure~\ref{fig:disorder}b shows centre transmission peak when randomly shifting the AER mechanical impedance values $\zeta$, directly related to $\omega_0$,  of every AER. In other words, the mass, resistance and compliance of each AER are gradually shifted from their nominal values effectively randomising their resonance frequencies using a uniform distribution. Surprisingly, remarkable resilience of the transmission peak can be observed up to disorders exceeding 40\% of the nominal impedance values $\zeta$. Indeed, the change of $\zeta$ doesn't significantly alter the scattering matrix far from the local resonance at the Bragg gap where topology occurs.
We argue that the symmetry responsible for the topological protection in this system is that of time-reversal symmetry and equality of the far-field scattering matrices as explained in a related study carried out by Zangeneh-Nejad \textit{et al.}~\cite{Zangeneh-Nejad2019} - albeit for a chain of passive scatterers. Their TMM-derived analytical model also produced an eigenvalue problem involving the single cell transfer matrix $M_{cell}$. Here, equality of the far-field scattering matrices of the AERs is obvious as both AERs composing the unit cell are assumed identical. Time-reversal symmetry on the other hand guarantees that $M_{cell}\in SU(1,1)$- the group of non-Hermitian matrices of the form:

\begin{equation}
	M_{cell} = \left( {\begin{array}{cc}
					\alpha & \beta \\
					\delta & \gamma \\
			\end{array} } \right)\in SU(1,1)  \Rightarrow 
	\begin{cases}
			&\alpha- \gamma^* = 0 \\
	 		&\beta - \delta^* = 0
	 	\end{cases}
\end{equation}

\begin{figure}[ht]
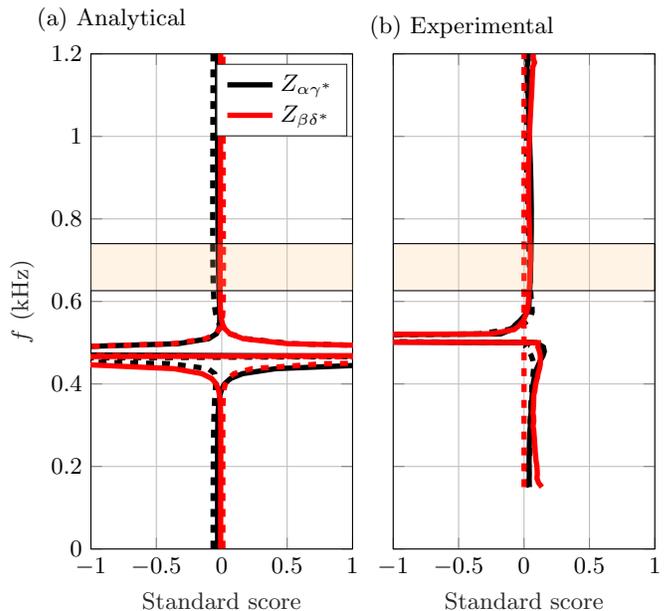

	\begin{minipage}[b]{0.21\textwidth}
		\begin{flushleft}
			\hspace{10pt} (a) Analytical
		\end{flushleft}
		\vspace{-3mm} 
		\input{./figures/tikz/su1_ana.tex}
	\end{minipage}
	\hfill
	\begin{minipage}[b]{0.21\textwidth}
		\begin{flushleft}
			(b) Experimental
		\end{flushleft}
		\vspace{-3mm} 
		\input{./figures/tikz/su1_exp.tex}
	\end{minipage}
	\caption{\label{fig:su1} Standard score $Z$ assessing the "$SU(1,1)$-ness" for the analytical model (a) and measured experimental data (b). Solid and dashed lines correspond to real and complex parts respectively. The largest Bragg gap from the dispersion of Figure~\ref{fig:disp}a is indicated by the orange transparent zones.}
\end{figure}

Keeping in mind that $M_{cell}$ is frequency-dependent, a measure of the "SU(1,1)-ness" of $M_{cell}$ can be assessed by computing the standard score $Z$ of the difference between the diagonal elements as a function of the angular frequency $\omega$:
 \begin{align}
	 	Z_{\alpha\gamma^*}(\omega) &= \frac{(\alpha(\omega)-\gamma^*(\omega)) - \mu_{\alpha\gamma^*}}{\sigma_{\alpha\gamma^*}}\\
	 	Z_{\beta\delta^*}(\omega) &= \frac{(\beta(\omega)-\delta^*(\omega)) - \mu_{\beta\delta^*}}{\sigma_{\beta\delta^*}}
	 \end{align}
where $\mu$ and $\sigma$ are the mean and standard deviations over frequency of the differences respectively. The latter is plotted for the maximum coupling $\delta = +a/4$ for the analytical and experimental single cell in Figures \ref{fig:su1}a and \ref{fig:su1}b respectively.

 The proximity to zero is a measure of the "SU(1,1)-ness" of $M_{cell}$. Around the locally resonant gap (highlighted in orange), the Z-score strongly diverges from the origin. This is due to the dissipative nature of the AERs which breaks time reversal symmetry - setting the mechanical resistance $R_m$ to zero in the model effectively neutralizes the normalized differences throughout the whole frequency range. At the Bragg gap where topology is relevant, the curves for both the analytical model and experimental data are much closer to the origin which indicates that the aforementioned topological theory can be applied.

\section{\label{sec:conc} Conclusions}

An active dimer composed of two electronically controlled AERs was presented. We demonstrated a TMM-derived analytical model and a current control scheme that enables shaping of both the locally resonant and the Bragg gaps. One active unit-cell was built and characterized by measuring its transfer matrix and subsequently deriving the dispersion curve by applying Bloch theorem. The proposed model yielded meta-crystal dispersion curves that closely matched those obtained from experiments. Most notably, we showed that the locally resonant gap and the Bragg gap could be easily shaped via speaker impedance synthesis and dimerisation parameter tuning $\delta$ respectively. With finite element modelling, we finally showed that a finite SSH-like meta-crystal composed of these active cells can host topological interface states protected by time-reversal symmetry and equality of the far-field scattering matrices. Owing to the wide tunability and reconfigurability, the presented active unit cell opens up a new path to seek out novel physical phenomena such as those involving time-modulated interaction, time reversal symmetry breaking or non-linearity. For example, the latter can be achieved by simple modification of the current control scheme  and  can thus serve as an experimental vessel for exploring non-linear topological systems \cite{Guo2020}.

\section*{Acknowledgements}
This research is supported by the Swiss National Science Foundation (SNSF) under grant No.~$200020\_200498$.

%

\appendix

\section{Dispersion measurement}\label{app:disp}
The dispersion of an acoustic meta-crystal can be obtained by measuring the transfer matrix of a single unit cell.
\begin{figure}[ht]
	\includegraphics[width=0.99\linewidth]{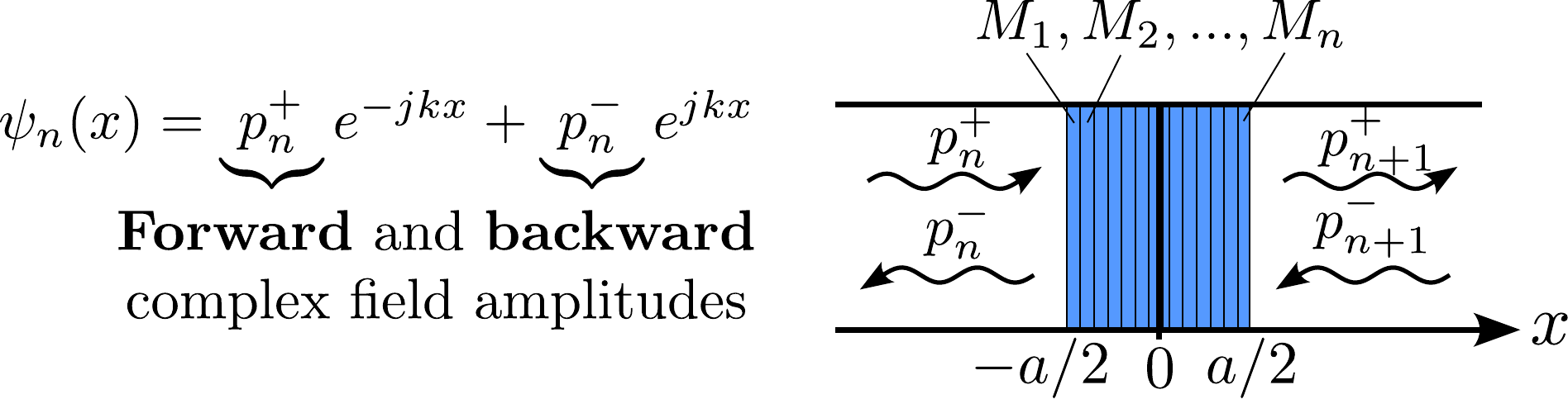}
	\caption{Transfer Matrix Method. The total cell transfer matrix $M_{cell}$ is composed of $n$ segment - each specified by it's own transfer matrix $M_n$}
	\label{fig:TMM}%
\end{figure}

Assuming continuity of pressure and volumetric flow, let there be a stratified medium of total length $a$ in an acoustic duct (Fig.~\ref{fig:TMM}). The total cell transfer matrix $M_{cell}$ relating the left and right complex field amplitudes is given by the matrix product of each sub-cell comprising the cell \cite{Richoux2002,Richoux2009,Wang2012}:
\begin{equation}
	M_{cell} = M_1 M_2 ... M_n
\end{equation}

Noting $\ket \psi  = (p_n^+,p_n^-)^T$, the application of the Bloch theorem with respect to the system periodicity yields the following Bloch Eigenvalue problem:

\begin{equation}
	M_{cell}(\omega)\ket \psi  =e^{j q_F(\omega)a} \ket \psi 
\end{equation}

where  $\omega$ is the angular frequency and $q_F$ is the Floquet-Bloch eigen-value. For each value of $\omega$, $M_{cell}(\omega)$ can be diagonalized which yields the band structure:
\begin{equation}
	q_F(\omega)= \arccos\left(\frac{1}{2} Tr(M_{cell}(\omega)) \right)/a
\end{equation}

The latter holds as long as reciprocity of the system is conserved \cite{Jimenez2021}, or equivalently $\det M = 1$.

\section{Impedance synthesis}\label{app:impedance_synth} 
This appendix summarizes the work reported by E. Rivet \textit{et al.} \cite{Rivet2017a}.
\begin{figure}[ht]
	\includegraphics[width=0.45\linewidth]{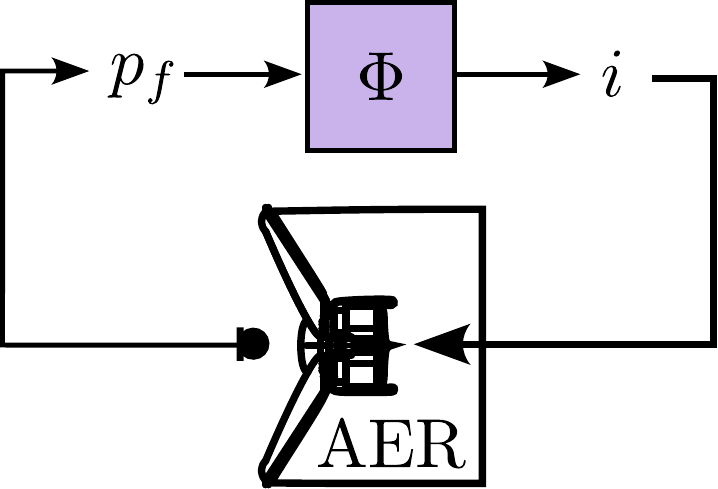}
	\caption{Impedance synthesis control scheme. The front pressure is measured at the front of the AER diaphragm (cross-section) with a microphone and is multiplied by a transfer function $\Phi$ which in turn outputs the drive current required to achieve a specified target impedance. }
	\label{fig:impedance_synth}%
\end{figure}
Let's consider a closed-box electrodynamic loudspeaker subject to an exogenous sound pressure $p_f$ at the diaphragm, and an electrical current $i$ at its electrical terminals as shown in figure~\ref{fig:impedance_synth}. Assuming steady-state and expressed in the frequency domain, its velocity response $v(s)$ is derived from Newton's second law of motion:

	\begin{equation}
		\zeta_{mc}(s)v(s) = S_d p_f(s) - B\ell i(s)
	\end{equation}
	
	Where:
	\begin{itemize}
		\item $\zeta_{mc}$: total mechanical impedance (N.s/m), including the compliance of air inside the cabinet
		\item $v$: diaphragm velocity (m/s)
		\item $S_d$: diaphragm surface area ($m^2$)
		\item $p_f$: pressure in front of diaphragm (Pa)
		\item $B\ell$: Force factor (T.m)
		\item $i$: electrical current (A)
		\item $s = j \omega$: the Laplace variable (Rad/s)
	\end{itemize}

	This allows deriving the electric current $i(s)$ required to achieve a target mechanical impedance $\zeta_{st}$ (Pa.s/m) at the diaphragm:
	
	\begin{equation}
		i(s) = \frac{(\zeta_{st}(s)-\zeta_{mc}(s))\cdot v(s)}{Bl}
	\end{equation}
	
	We can specify 	$\zeta_{st}$ to have the following form:
	
	\begin{equation}
		\zeta_{st}(s) = \mu_M M_{ms}\cdot s+\mu_R R_{ms} + \mu_C/(C_{mc}\cdot s)
	\end{equation}
	
	Where:
	\begin{itemize}
		\item $M_{ms}$: diaphragm mass (kg)
		\item $R_{ms}$: mechanical resistance  
		\item $C_{mc}$: speaker + cabinet compliance ($s^2$/kg)
		\item $\mu_M,\mu_R,\mu_C$: control parameters
	\end{itemize}
	Thus, the control current required to synthesize a target virtual impedance $\zeta_{st}$ is:
	
	\begin{equation}
		i(s) = p_f(s)\cdot\Phi(s)
	\end{equation}
	Where
	\begin{widetext}
		\begin{equation} 
			\Phi = \frac{S_d}{Bl}\left[\frac{(\mu_M-1)M_{ms}\cdot s+(\mu_R-1)R_{ms} + (\mu_C-1)/(C_{mc}\cdot s)}{\mu_M M_{ms}\cdot s+\mu_R R_{ms} + 	\mu_C/(C_{mc}\cdot s)}\right]
		\end{equation}
	\end{widetext}
	
	is the transfer function.
	
	 In practice, sound waves arriving at the microphones are detected, digitally sampled, and processed in real-time by a software program running on a commercial Performance Speedgoat machine equipped with the IO334 module. The digital output signals are then converted back to analog and are fed back to the speakers to generate the desired synthesized response. The Speedgoat Performance model can reliably deliver processing delays as little as 25 $\mu s$ which enables large synthesis as explained in a recent stability study carried out by De Bono \textit{et al.} \cite{DeBono2022}.

%
%

\bibliography{library}

\begin{thebibliography}{44}%
\makeatletter
\providecommand \@ifxundefined [1]{%
 \@ifx{#1\undefined}
}%
\providecommand \@ifnum [1]{%
 \ifnum #1\expandafter \@firstoftwo
 \else \expandafter \@secondoftwo
 \fi
}%
\providecommand \@ifx [1]{%
 \ifx #1\expandafter \@firstoftwo
 \else \expandafter \@secondoftwo
 \fi
}%
\providecommand \natexlab [1]{#1}%
\providecommand \enquote  [1]{``#1''}%
\providecommand \bibnamefont  [1]{#1}%
\providecommand \bibfnamefont [1]{#1}%
\providecommand \citenamefont [1]{#1}%
\providecommand \href@noop [0]{\@secondoftwo}%
\providecommand \href [0]{\begingroup \@sanitize@url \@href}%
\providecommand \@href[1]{\@@startlink{#1}\@@href}%
\providecommand \@@href[1]{\endgroup#1\@@endlink}%
\providecommand \@sanitize@url [0]{\catcode `\\12\catcode `\$12\catcode
  `\&12\catcode `\#12\catcode `\^12\catcode `\_12\catcode `\%12\relax}%
\providecommand \@@startlink[1]{}%
\providecommand \@@endlink[0]{}%
\providecommand \url  [0]{\begingroup\@sanitize@url \@url }%
\providecommand \@url [1]{\endgroup\@href {#1}{\urlprefix }}%
\providecommand \urlprefix  [0]{URL }%
\providecommand \Eprint [0]{\href }%
\providecommand \doibase [0]{https://doi.org/}%
\providecommand \selectlanguage [0]{\@gobble}%
\providecommand \bibinfo  [0]{\@secondoftwo}%
\providecommand \bibfield  [0]{\@secondoftwo}%
\providecommand \translation [1]{[#1]}%
\providecommand \BibitemOpen [0]{}%
\providecommand \bibitemStop [0]{}%
\providecommand \bibitemNoStop [0]{.\EOS\space}%
\providecommand \EOS [0]{\spacefactor3000\relax}%
\providecommand \BibitemShut  [1]{\csname bibitem#1\endcsname}%
\let\auto@bib@innerbib\@empty
\bibitem [{\citenamefont {He}\ and\ \citenamefont {Maynard}(1986)}]{He1986}%
  \BibitemOpen
  \bibfield  {author} {\bibinfo {author} {\bibfnamefont {S.}~\bibnamefont
  {He}}\ and\ \bibinfo {author} {\bibfnamefont {J.~D.}\ \bibnamefont
  {Maynard}},\ }\bibfield  {title} {\bibinfo {title} {{Detailed measurements of
  inelastic scattering in Anderson localization}},\ }\href
  {https://doi.org/10.1103/PhysRevLett.57.3171} {\bibfield  {journal} {\bibinfo
   {journal} {Physical Review Letters}\ }\textbf {\bibinfo {volume} {57}},\
  \bibinfo {pages} {3171} (\bibinfo {year} {1986})}\BibitemShut {NoStop}%
\bibitem [{\citenamefont {Maynard}(1992)}]{Maynard1992}%
  \BibitemOpen
  \bibfield  {author} {\bibinfo {author} {\bibfnamefont {J.~D.}\ \bibnamefont
  {Maynard}},\ }\bibfield  {title} {\bibinfo {title} {{A possible explanation
  for the discrepancy in electron persistent current amplitudes: A superfluid
  persistent current analog}},\ }\href {https://doi.org/10.1007/BF00692587}
  {\bibfield  {journal} {\bibinfo  {journal} {Journal of Low Temperature
  Physics}\ }\textbf {\bibinfo {volume} {89}},\ \bibinfo {pages} {155}
  (\bibinfo {year} {1992})}\BibitemShut {NoStop}%
\bibitem [{\citenamefont {Xiao}\ \emph {et~al.}(2015)\citenamefont {Xiao},
  \citenamefont {Ma}, \citenamefont {Yang}, \citenamefont {Sheng},
  \citenamefont {Zhang},\ and\ \citenamefont {Chan}}]{Xiao2015}%
  \BibitemOpen
  \bibfield  {author} {\bibinfo {author} {\bibfnamefont {M.}~\bibnamefont
  {Xiao}}, \bibinfo {author} {\bibfnamefont {G.}~\bibnamefont {Ma}}, \bibinfo
  {author} {\bibfnamefont {Z.}~\bibnamefont {Yang}}, \bibinfo {author}
  {\bibfnamefont {P.}~\bibnamefont {Sheng}}, \bibinfo {author} {\bibfnamefont
  {Z.~Q.}\ \bibnamefont {Zhang}},\ and\ \bibinfo {author} {\bibfnamefont
  {C.~T.}\ \bibnamefont {Chan}},\ }\bibfield  {title} {\bibinfo {title}
  {{Geometric phase and band inversion in periodic acoustic systems}},\ }\href
  {https://doi.org/10.1038/nphys3228} {\bibfield  {journal} {\bibinfo
  {journal} {Nature Physics}\ }\textbf {\bibinfo {volume} {11}},\ \bibinfo
  {pages} {240} (\bibinfo {year} {2015})}\BibitemShut {NoStop}%
\bibitem [{\citenamefont {Zangeneh-Nejad}\ and\ \citenamefont
  {Fleury}(2019)}]{Zangeneh-Nejad2019}%
  \BibitemOpen
  \bibfield  {author} {\bibinfo {author} {\bibfnamefont {F.}~\bibnamefont
  {Zangeneh-Nejad}}\ and\ \bibinfo {author} {\bibfnamefont {R.}~\bibnamefont
  {Fleury}},\ }\bibfield  {title} {\bibinfo {title} {{Topological analog signal
  processing}},\ }\href {https://doi.org/10.1038/s41467-019-10086-3} {\bibfield
   {journal} {\bibinfo  {journal} {Nature Communications}\ }\textbf {\bibinfo
  {volume} {10}},\ \bibinfo {pages} {1} (\bibinfo {year} {2019})}\BibitemShut
  {NoStop}%
\bibitem [{\citenamefont {Yang}\ and\ \citenamefont {Zhang}(2016)}]{Yang2016}%
  \BibitemOpen
  \bibfield  {author} {\bibinfo {author} {\bibfnamefont {Z.}~\bibnamefont
  {Yang}}\ and\ \bibinfo {author} {\bibfnamefont {B.}~\bibnamefont {Zhang}},\
  }\bibfield  {title} {\bibinfo {title} {{Acoustic Type-II Weyl Nodes from
  Stacking Dimerized Chains}},\ }\href
  {https://doi.org/10.1103/PhysRevLett.117.224301} {\bibfield  {journal}
  {\bibinfo  {journal} {Physical Review Letters}\ }\textbf {\bibinfo {volume}
  {117}},\ \bibinfo {pages} {1} (\bibinfo {year} {2016})}\BibitemShut {NoStop}%
\bibitem [{\citenamefont {Esmann}\ \emph {et~al.}(2018)\citenamefont {Esmann},
  \citenamefont {Lamberti}, \citenamefont {Lema{\^{i}}tre},\ and\ \citenamefont
  {Lanzillotti-Kimura}}]{Esmann2018}%
  \BibitemOpen
  \bibfield  {author} {\bibinfo {author} {\bibfnamefont {M.}~\bibnamefont
  {Esmann}}, \bibinfo {author} {\bibfnamefont {F.~R.}\ \bibnamefont
  {Lamberti}}, \bibinfo {author} {\bibfnamefont {A.}~\bibnamefont
  {Lema{\^{i}}tre}},\ and\ \bibinfo {author} {\bibfnamefont {N.~D.}\
  \bibnamefont {Lanzillotti-Kimura}},\ }\bibfield  {title} {\bibinfo {title}
  {{Topological acoustics in coupled nanocavity arrays}},\ }\href
  {https://doi.org/10.1103/PhysRevB.98.161109} {\bibfield  {journal} {\bibinfo
  {journal} {Physical Review B}\ }\textbf {\bibinfo {volume} {98}},\ \bibinfo
  {pages} {161109} (\bibinfo {year} {2018})}\BibitemShut {NoStop}%
\bibitem [{\citenamefont {Shen}\ \emph {et~al.}(2020)\citenamefont {Shen},
  \citenamefont {Zeng}, \citenamefont {Geng}, \citenamefont {Zhao},
  \citenamefont {Peng},\ and\ \citenamefont {Zhu}}]{Shen2020}%
  \BibitemOpen
  \bibfield  {author} {\bibinfo {author} {\bibfnamefont {Y.~X.}\ \bibnamefont
  {Shen}}, \bibinfo {author} {\bibfnamefont {L.~S.}\ \bibnamefont {Zeng}},
  \bibinfo {author} {\bibfnamefont {Z.~G.}\ \bibnamefont {Geng}}, \bibinfo
  {author} {\bibfnamefont {D.~G.}\ \bibnamefont {Zhao}}, \bibinfo {author}
  {\bibfnamefont {Y.~G.}\ \bibnamefont {Peng}},\ and\ \bibinfo {author}
  {\bibfnamefont {X.~F.}\ \bibnamefont {Zhu}},\ }\bibfield  {title} {\bibinfo
  {title} {{Acoustic Adiabatic Propagation Based on Topological Pumping in a
  Coupled Multicavity Chain Lattice}},\ }\href
  {https://doi.org/10.1103/PhysRevApplied.14.014043} {\bibfield  {journal}
  {\bibinfo  {journal} {Physical Review Applied}\ }\textbf {\bibinfo {volume}
  {14}},\ \bibinfo {pages} {1} (\bibinfo {year} {2020})}\BibitemShut {NoStop}%
\bibitem [{\citenamefont {Yan}\ \emph {et~al.}(2020)\citenamefont {Yan},
  \citenamefont {Huang}, \citenamefont {Luo}, \citenamefont {Lu}, \citenamefont
  {Deng},\ and\ \citenamefont {Liu}}]{Yan2020}%
  \BibitemOpen
  \bibfield  {author} {\bibinfo {author} {\bibfnamefont {M.}~\bibnamefont
  {Yan}}, \bibinfo {author} {\bibfnamefont {X.}~\bibnamefont {Huang}}, \bibinfo
  {author} {\bibfnamefont {L.}~\bibnamefont {Luo}}, \bibinfo {author}
  {\bibfnamefont {J.}~\bibnamefont {Lu}}, \bibinfo {author} {\bibfnamefont
  {W.}~\bibnamefont {Deng}},\ and\ \bibinfo {author} {\bibfnamefont
  {Z.}~\bibnamefont {Liu}},\ }\bibfield  {title} {\bibinfo {title} {{Acoustic
  square-root topological states}},\ }\href
  {https://doi.org/10.1103/PhysRevB.102.180102} {\bibfield  {journal} {\bibinfo
   {journal} {Physical Review B}\ }\textbf {\bibinfo {volume} {102}},\ \bibinfo
  {pages} {180102} (\bibinfo {year} {2020})}\BibitemShut {NoStop}%
\bibitem [{\citenamefont {Chen}\ \emph {et~al.}(2020)\citenamefont {Chen},
  \citenamefont {Wang}, \citenamefont {Zhang},\ and\ \citenamefont
  {Ma}}]{Chen2020}%
  \BibitemOpen
  \bibfield  {author} {\bibinfo {author} {\bibfnamefont {Z.~G.}\ \bibnamefont
  {Chen}}, \bibinfo {author} {\bibfnamefont {L.}~\bibnamefont {Wang}}, \bibinfo
  {author} {\bibfnamefont {G.}~\bibnamefont {Zhang}},\ and\ \bibinfo {author}
  {\bibfnamefont {G.}~\bibnamefont {Ma}},\ }\bibfield  {title} {\bibinfo
  {title} {{Chiral symmetry breaking of tight-binding models in coupled
  acoustic-cavity systems}},\ }\href
  {https://doi.org/10.1103/PhysRevApplied.14.024023} {\bibfield  {journal}
  {\bibinfo  {journal} {Physical Review Applied}\ }\textbf {\bibinfo {volume}
  {14}},\ \bibinfo {pages} {1} (\bibinfo {year} {2020})}\BibitemShut {NoStop}%
\bibitem [{\citenamefont {Zhao}\ \emph {et~al.}(2021)\citenamefont {Zhao},
  \citenamefont {Chen}, \citenamefont {Li},\ and\ \citenamefont
  {Zhu}}]{Zhao2021}%
  \BibitemOpen
  \bibfield  {author} {\bibinfo {author} {\bibfnamefont {D.}~\bibnamefont
  {Zhao}}, \bibinfo {author} {\bibfnamefont {X.}~\bibnamefont {Chen}}, \bibinfo
  {author} {\bibfnamefont {P.}~\bibnamefont {Li}},\ and\ \bibinfo {author}
  {\bibfnamefont {X.~F.}\ \bibnamefont {Zhu}},\ }\bibfield  {title} {\bibinfo
  {title} {{Subwavelength acoustic energy harvesting via topological interface
  states in 1D Helmholtz resonator arrays}},\ }\href
  {https://doi.org/10.1063/5.0034811} {\bibfield  {journal} {\bibinfo
  {journal} {AIP Advances}\ }\textbf {\bibinfo {volume} {11}},\ \bibinfo
  {pages} {015241} (\bibinfo {year} {2021})}\BibitemShut {NoStop}%
\bibitem [{\citenamefont {Aur{\'{e}}gan}\ and\ \citenamefont
  {Pagneux}(2017)}]{Auregan2017}%
  \BibitemOpen
  \bibfield  {author} {\bibinfo {author} {\bibfnamefont {Y.}~\bibnamefont
  {Aur{\'{e}}gan}}\ and\ \bibinfo {author} {\bibfnamefont {V.}~\bibnamefont
  {Pagneux}},\ }\bibfield  {title} {\bibinfo {title} {{P T -Symmetric
  Scattering in Flow Duct Acoustics}},\ }\href
  {https://doi.org/10.1103/PhysRevLett.118.174301} {\bibfield  {journal}
  {\bibinfo  {journal} {Physical Review Letters}\ }\textbf {\bibinfo {volume}
  {118}},\ \bibinfo {pages} {174301} (\bibinfo {year} {2017})},\ \Eprint
  {https://arxiv.org/abs/1701.07618} {arXiv:1701.07618} \BibitemShut {NoStop}%
\bibitem [{\citenamefont {Christensen}\ \emph {et~al.}(2016)\citenamefont
  {Christensen}, \citenamefont {Willatzen}, \citenamefont {Velasco},\ and\
  \citenamefont {Lu}}]{Christensen2016}%
  \BibitemOpen
  \bibfield  {author} {\bibinfo {author} {\bibfnamefont {J.}~\bibnamefont
  {Christensen}}, \bibinfo {author} {\bibfnamefont {M.}~\bibnamefont
  {Willatzen}}, \bibinfo {author} {\bibfnamefont {V.~R.}\ \bibnamefont
  {Velasco}},\ and\ \bibinfo {author} {\bibfnamefont {M.~H.}\ \bibnamefont
  {Lu}},\ }\bibfield  {title} {\bibinfo {title} {{Parity-Time Synthetic
  Phononic Media}},\ }\href
  {https://doi.org/10.1103/PHYSREVLETT.116.207601/FIGURES/4/MEDIUM} {\bibfield
  {journal} {\bibinfo  {journal} {Physical Review Letters}\ }\textbf {\bibinfo
  {volume} {116}},\ \bibinfo {pages} {207601} (\bibinfo {year}
  {2016})}\BibitemShut {NoStop}%
\bibitem [{\citenamefont {Fleury}\ \emph
  {et~al.}(2015{\natexlab{a}})\citenamefont {Fleury}, \citenamefont {Sounas},\
  and\ \citenamefont {Al{\`{u}}}}]{Fleury2015}%
  \BibitemOpen
  \bibfield  {author} {\bibinfo {author} {\bibfnamefont {R.}~\bibnamefont
  {Fleury}}, \bibinfo {author} {\bibfnamefont {D.}~\bibnamefont {Sounas}},\
  and\ \bibinfo {author} {\bibfnamefont {A.}~\bibnamefont {Al{\`{u}}}},\
  }\bibfield  {title} {\bibinfo {title} {{An invisible acoustic sensor based on
  parity-time symmetry}},\ }\href {https://doi.org/10.1038/ncomms6905}
  {\bibfield  {journal} {\bibinfo  {journal} {Nature Communications}\ }\textbf
  {\bibinfo {volume} {6}},\ \bibinfo {pages} {1} (\bibinfo {year}
  {2015}{\natexlab{a}})}\BibitemShut {NoStop}%
\bibitem [{\citenamefont {Shi}\ \emph {et~al.}(2016)\citenamefont {Shi},
  \citenamefont {Dubois}, \citenamefont {Chen}, \citenamefont {Cheng},
  \citenamefont {Ramezani}, \citenamefont {Wang},\ and\ \citenamefont
  {Zhang}}]{Shi2016}%
  \BibitemOpen
  \bibfield  {author} {\bibinfo {author} {\bibfnamefont {C.}~\bibnamefont
  {Shi}}, \bibinfo {author} {\bibfnamefont {M.}~\bibnamefont {Dubois}},
  \bibinfo {author} {\bibfnamefont {Y.}~\bibnamefont {Chen}}, \bibinfo {author}
  {\bibfnamefont {L.}~\bibnamefont {Cheng}}, \bibinfo {author} {\bibfnamefont
  {H.}~\bibnamefont {Ramezani}}, \bibinfo {author} {\bibfnamefont
  {Y.}~\bibnamefont {Wang}},\ and\ \bibinfo {author} {\bibfnamefont
  {X.}~\bibnamefont {Zhang}},\ }\bibfield  {title} {\bibinfo {title}
  {{Accessing the exceptional points of parity-time symmetric acoustics}},\
  }\href {https://doi.org/10.1038/ncomms11110} {\bibfield  {journal} {\bibinfo
  {journal} {Nature Communications}\ }\textbf {\bibinfo {volume} {7}},\
  \bibinfo {pages} {1} (\bibinfo {year} {2016})}\BibitemShut {NoStop}%
\bibitem [{\citenamefont {Zhang}\ \emph {et~al.}(2021)\citenamefont {Zhang},
  \citenamefont {Yang}, \citenamefont {Ge}, \citenamefont {Guan}, \citenamefont
  {Chen}, \citenamefont {Yan}, \citenamefont {Chen}, \citenamefont {Xi},
  \citenamefont {Li}, \citenamefont {Jia}, \citenamefont {Yuan}, \citenamefont
  {Sun}, \citenamefont {Chen},\ and\ \citenamefont {Zhang}}]{Zhang2021}%
  \BibitemOpen
  \bibfield  {author} {\bibinfo {author} {\bibfnamefont {L.}~\bibnamefont
  {Zhang}}, \bibinfo {author} {\bibfnamefont {Y.}~\bibnamefont {Yang}},
  \bibinfo {author} {\bibfnamefont {Y.}~\bibnamefont {Ge}}, \bibinfo {author}
  {\bibfnamefont {Y.~J.}\ \bibnamefont {Guan}}, \bibinfo {author}
  {\bibfnamefont {Q.}~\bibnamefont {Chen}}, \bibinfo {author} {\bibfnamefont
  {Q.}~\bibnamefont {Yan}}, \bibinfo {author} {\bibfnamefont {F.}~\bibnamefont
  {Chen}}, \bibinfo {author} {\bibfnamefont {R.}~\bibnamefont {Xi}}, \bibinfo
  {author} {\bibfnamefont {Y.}~\bibnamefont {Li}}, \bibinfo {author}
  {\bibfnamefont {D.}~\bibnamefont {Jia}}, \bibinfo {author} {\bibfnamefont
  {S.~Q.}\ \bibnamefont {Yuan}}, \bibinfo {author} {\bibfnamefont {H.~X.}\
  \bibnamefont {Sun}}, \bibinfo {author} {\bibfnamefont {H.}~\bibnamefont
  {Chen}},\ and\ \bibinfo {author} {\bibfnamefont {B.}~\bibnamefont {Zhang}},\
  }\bibfield  {title} {\bibinfo {title} {{Acoustic non-Hermitian skin effect
  from twisted winding topology}},\ }\href
  {https://doi.org/10.1038/s41467-021-26619-8} {\bibfield  {journal} {\bibinfo
  {journal} {Nature Communications}\ }\textbf {\bibinfo {volume} {12}},\
  \bibinfo {pages} {6} (\bibinfo {year} {2021})}\BibitemShut {NoStop}%
\bibitem [{\citenamefont {Liu}\ \emph {et~al.}(2022)\citenamefont {Liu},
  \citenamefont {Li}, \citenamefont {Chen}, \citenamefont {Tang}, \citenamefont
  {Chen}, \citenamefont {Liang}, \citenamefont {Ma},\ and\ \citenamefont
  {Cheng}}]{Liu2022}%
  \BibitemOpen
  \bibfield  {author} {\bibinfo {author} {\bibfnamefont {J.~J.}\ \bibnamefont
  {Liu}}, \bibinfo {author} {\bibfnamefont {Z.~W.}\ \bibnamefont {Li}},
  \bibinfo {author} {\bibfnamefont {Z.~G.}\ \bibnamefont {Chen}}, \bibinfo
  {author} {\bibfnamefont {W.}~\bibnamefont {Tang}}, \bibinfo {author}
  {\bibfnamefont {A.}~\bibnamefont {Chen}}, \bibinfo {author} {\bibfnamefont
  {B.}~\bibnamefont {Liang}}, \bibinfo {author} {\bibfnamefont
  {G.}~\bibnamefont {Ma}},\ and\ \bibinfo {author} {\bibfnamefont {J.~C.}\
  \bibnamefont {Cheng}},\ }\bibfield  {title} {\bibinfo {title} {{Experimental
  Realization of Weyl Exceptional Rings in a Synthetic Three-Dimensional
  Non-Hermitian Phononic Crystal}},\ }\href
  {https://doi.org/10.1103/PhysRevLett.129.084301} {\bibfield  {journal}
  {\bibinfo  {journal} {Physical Review Letters}\ }\textbf {\bibinfo {volume}
  {129}},\ \bibinfo {pages} {084301} (\bibinfo {year} {2022})}\BibitemShut
  {NoStop}%
\bibitem [{\citenamefont {Ma}\ and\ \citenamefont {Sheng}(2016)}]{Ma2016}%
  \BibitemOpen
  \bibfield  {author} {\bibinfo {author} {\bibfnamefont {G.}~\bibnamefont
  {Ma}}\ and\ \bibinfo {author} {\bibfnamefont {P.}~\bibnamefont {Sheng}},\
  }\bibfield  {title} {\bibinfo {title} {{Acoustic metamaterials: From local
  resonances to broad horizons}},\ }\bibfield  {journal} {\bibinfo  {journal}
  {Science Advances}\ }\textbf {\bibinfo {volume} {2}},\ \href
  {https://doi.org/10.1126/sciadv.1501595} {10.1126/sciadv.1501595} (\bibinfo
  {year} {2016})\BibitemShut {NoStop}%
\bibitem [{\citenamefont {Popa}\ and\ \citenamefont {Cummer}(2014)}]{Popa2014}%
  \BibitemOpen
  \bibfield  {author} {\bibinfo {author} {\bibfnamefont {B.~I.}\ \bibnamefont
  {Popa}}\ and\ \bibinfo {author} {\bibfnamefont {S.~A.}\ \bibnamefont
  {Cummer}},\ }\bibfield  {title} {\bibinfo {title} {{Non-reciprocal and highly
  nonlinear active acoustic metamaterials}},\ }\href
  {https://doi.org/10.1038/ncomms4398} {\bibfield  {journal} {\bibinfo
  {journal} {Nature communications}\ }\textbf {\bibinfo {volume} {5}},\
  \bibinfo {pages} {3398} (\bibinfo {year} {2014})}\BibitemShut {NoStop}%
\bibitem [{\citenamefont {Boechler}\ \emph {et~al.}(2011)\citenamefont
  {Boechler}, \citenamefont {Theocharis},\ and\ \citenamefont
  {Daraio}}]{Boechler2011}%
  \BibitemOpen
  \bibfield  {author} {\bibinfo {author} {\bibfnamefont {N.}~\bibnamefont
  {Boechler}}, \bibinfo {author} {\bibfnamefont {G.}~\bibnamefont
  {Theocharis}},\ and\ \bibinfo {author} {\bibfnamefont {C.}~\bibnamefont
  {Daraio}},\ }\bibfield  {title} {\bibinfo {title} {{Bifurcation-based
  acoustic switching and rectification}},\ }\href
  {https://doi.org/10.1038/nmat3072} {\bibfield  {journal} {\bibinfo  {journal}
  {Nature Materials}\ }\textbf {\bibinfo {volume} {10}},\ \bibinfo {pages}
  {665} (\bibinfo {year} {2011})}\BibitemShut {NoStop}%
\bibitem [{\citenamefont {Zangeneh-Nejad}\ and\ \citenamefont
  {Fleury}(2018)}]{Zangeneh-nejad2018}%
  \BibitemOpen
  \bibfield  {author} {\bibinfo {author} {\bibfnamefont {F.}~\bibnamefont
  {Zangeneh-Nejad}}\ and\ \bibinfo {author} {\bibfnamefont {R.}~\bibnamefont
  {Fleury}},\ }\bibfield  {title} {\bibinfo {title} {{Doppler-based acoustic
  gyrator}},\ }\href {https://doi.org/10.3390/app8071083} {\bibfield  {journal}
  {\bibinfo  {journal} {Applied Sciences (Switzerland)}\ }\textbf {\bibinfo
  {volume} {8}},\ \bibinfo {pages} {1083} (\bibinfo {year} {2018})}\BibitemShut
  {NoStop}%
\bibitem [{\citenamefont {Liang}\ \emph {et~al.}(2010)\citenamefont {Liang},
  \citenamefont {Guo}, \citenamefont {Tu}, \citenamefont {Zhang},\ and\
  \citenamefont {Cheng}}]{Liang2010}%
  \BibitemOpen
  \bibfield  {author} {\bibinfo {author} {\bibfnamefont {B.}~\bibnamefont
  {Liang}}, \bibinfo {author} {\bibfnamefont {X.~S.}\ \bibnamefont {Guo}},
  \bibinfo {author} {\bibfnamefont {J.}~\bibnamefont {Tu}}, \bibinfo {author}
  {\bibfnamefont {D.}~\bibnamefont {Zhang}},\ and\ \bibinfo {author}
  {\bibfnamefont {J.~C.}\ \bibnamefont {Cheng}},\ }\bibfield  {title} {\bibinfo
  {title} {{An acoustic rectifier}},\ }\href {https://doi.org/10.1038/nmat2881}
  {\bibfield  {journal} {\bibinfo  {journal} {Nature Materials}\ }\textbf
  {\bibinfo {volume} {9}},\ \bibinfo {pages} {989} (\bibinfo {year}
  {2010})}\BibitemShut {NoStop}%
\bibitem [{\citenamefont {Fleury}\ \emph {et~al.}(2014)\citenamefont {Fleury},
  \citenamefont {Sounas}, \citenamefont {Sieck}, \citenamefont {Haberman},\
  and\ \citenamefont {Al{\`{u}}}}]{Fleury2014}%
  \BibitemOpen
  \bibfield  {author} {\bibinfo {author} {\bibfnamefont {R.}~\bibnamefont
  {Fleury}}, \bibinfo {author} {\bibfnamefont {D.~L.}\ \bibnamefont {Sounas}},
  \bibinfo {author} {\bibfnamefont {C.~F.}\ \bibnamefont {Sieck}}, \bibinfo
  {author} {\bibfnamefont {M.~R.}\ \bibnamefont {Haberman}},\ and\ \bibinfo
  {author} {\bibfnamefont {A.}~\bibnamefont {Al{\`{u}}}},\ }\bibfield  {title}
  {\bibinfo {title} {{Sound isolation and giant linear nonreciprocity in a
  compact acoustic circulator}},\ }\href
  {https://doi.org/10.1126/science.1246957} {\bibfield  {journal} {\bibinfo
  {journal} {Science}\ }\textbf {\bibinfo {volume} {343}},\ \bibinfo {pages}
  {516} (\bibinfo {year} {2014})}\BibitemShut {NoStop}%
\bibitem [{\citenamefont {Fleury}\ \emph
  {et~al.}(2015{\natexlab{b}})\citenamefont {Fleury}, \citenamefont {Sounas},\
  and\ \citenamefont {Al{\`{u}}}}]{Fleury2015c}%
  \BibitemOpen
  \bibfield  {author} {\bibinfo {author} {\bibfnamefont {R.}~\bibnamefont
  {Fleury}}, \bibinfo {author} {\bibfnamefont {D.~L.}\ \bibnamefont {Sounas}},\
  and\ \bibinfo {author} {\bibfnamefont {A.}~\bibnamefont {Al{\`{u}}}},\
  }\bibfield  {title} {\bibinfo {title} {{Subwavelength ultrasonic circulator
  based on spatiotemporal modulation}},\ }\href
  {https://doi.org/10.1103/PHYSREVB.91.174306/FIGURES/5/MEDIUM} {\bibfield
  {journal} {\bibinfo  {journal} {Physical Review B - Condensed Matter and
  Materials Physics}\ }\textbf {\bibinfo {volume} {91}},\ \bibinfo {pages}
  {174306} (\bibinfo {year} {2015}{\natexlab{b}})}\BibitemShut {NoStop}%
\bibitem [{\citenamefont {Popa}\ \emph {et~al.}(2015)\citenamefont {Popa},
  \citenamefont {Shinde}, \citenamefont {Konneker},\ and\ \citenamefont
  {Cummer}}]{Popa2015}%
  \BibitemOpen
  \bibfield  {author} {\bibinfo {author} {\bibfnamefont {B.~I.}\ \bibnamefont
  {Popa}}, \bibinfo {author} {\bibfnamefont {D.}~\bibnamefont {Shinde}},
  \bibinfo {author} {\bibfnamefont {A.}~\bibnamefont {Konneker}},\ and\
  \bibinfo {author} {\bibfnamefont {S.~A.}\ \bibnamefont {Cummer}},\ }\bibfield
   {title} {\bibinfo {title} {{Active acoustic metamaterials reconfigurable in
  real time}},\ }\href {https://doi.org/10.1103/PhysRevB.91.220303} {\bibfield
  {journal} {\bibinfo  {journal} {Physical Review B - Condensed Matter and
  Materials Physics}\ }\textbf {\bibinfo {volume} {91}},\ \bibinfo {pages}
  {220303} (\bibinfo {year} {2015})},\ \Eprint
  {https://arxiv.org/abs/1505.00453} {arXiv:1505.00453} \BibitemShut {NoStop}%
\bibitem [{\citenamefont {Cho}\ \emph {et~al.}(2020)\citenamefont {Cho},
  \citenamefont {Wen}, \citenamefont {Park},\ and\ \citenamefont
  {Li}}]{Cho2020}%
  \BibitemOpen
  \bibfield  {author} {\bibinfo {author} {\bibfnamefont {C.}~\bibnamefont
  {Cho}}, \bibinfo {author} {\bibfnamefont {X.}~\bibnamefont {Wen}}, \bibinfo
  {author} {\bibfnamefont {N.}~\bibnamefont {Park}},\ and\ \bibinfo {author}
  {\bibfnamefont {J.}~\bibnamefont {Li}},\ }\bibfield  {title} {\bibinfo
  {title} {{Digitally virtualized atoms for acoustic metamaterials}},\ }\href
  {https://doi.org/10.1038/s41467-019-14124-y} {\bibfield  {journal} {\bibinfo
  {journal} {Nature Communications}\ }\textbf {\bibinfo {volume} {11}},\
  \bibinfo {pages} {1} (\bibinfo {year} {2020})}\BibitemShut {NoStop}%
\bibitem [{\citenamefont {Rivet}\ \emph
  {et~al.}(2017{\natexlab{a}})\citenamefont {Rivet}, \citenamefont {Karkar},\
  and\ \citenamefont {Lissek}}]{Rivet2017a}%
  \BibitemOpen
  \bibfield  {author} {\bibinfo {author} {\bibfnamefont {E.}~\bibnamefont
  {Rivet}}, \bibinfo {author} {\bibfnamefont {S.}~\bibnamefont {Karkar}},\ and\
  \bibinfo {author} {\bibfnamefont {H.}~\bibnamefont {Lissek}},\ }\bibfield
  {title} {\bibinfo {title} {{On the optimisation of multi-degree-of-freedom
  acoustic impedances of low-frequency electroacoustic absorbers for room modal
  equalisation}},\ }\href {https://doi.org/10.3813/AAA.919132} {\bibfield
  {journal} {\bibinfo  {journal} {Acta Acustica united with Acustica}\ }\textbf
  {\bibinfo {volume} {103}},\ \bibinfo {pages} {1025} (\bibinfo {year}
  {2017}{\natexlab{a}})}\BibitemShut {NoStop}%
\bibitem [{\citenamefont {Rivet}\ \emph
  {et~al.}(2017{\natexlab{b}})\citenamefont {Rivet}, \citenamefont {Karkar},\
  and\ \citenamefont {Lissek}}]{Rivet2017}%
  \BibitemOpen
  \bibfield  {author} {\bibinfo {author} {\bibfnamefont {E.}~\bibnamefont
  {Rivet}}, \bibinfo {author} {\bibfnamefont {S.}~\bibnamefont {Karkar}},\ and\
  \bibinfo {author} {\bibfnamefont {H.}~\bibnamefont {Lissek}},\ }\bibfield
  {title} {\bibinfo {title} {{On the optimisation of multi-degree-of-freedom
  acoustic impedances of low-frequency electroacoustic absorbers for room modal
  equalisation}},\ }\href {https://doi.org/10.3813/AAA.919132} {\bibfield
  {journal} {\bibinfo  {journal} {Acta Acustica united with Acustica}\ }\textbf
  {\bibinfo {volume} {103}},\ \bibinfo {pages} {1025} (\bibinfo {year}
  {2017}{\natexlab{b}})}\BibitemShut {NoStop}%
\bibitem [{\citenamefont {Boulandet}\ \emph {et~al.}(2018)\citenamefont
  {Boulandet}, \citenamefont {Lissek}, \citenamefont {Karkar}, \citenamefont
  {Collet}, \citenamefont {Matten}, \citenamefont {Ouisse},\ and\ \citenamefont
  {Versaevel}}]{Boulandet2018}%
  \BibitemOpen
  \bibfield  {author} {\bibinfo {author} {\bibfnamefont {R.}~\bibnamefont
  {Boulandet}}, \bibinfo {author} {\bibfnamefont {H.}~\bibnamefont {Lissek}},
  \bibinfo {author} {\bibfnamefont {S.}~\bibnamefont {Karkar}}, \bibinfo
  {author} {\bibfnamefont {M.}~\bibnamefont {Collet}}, \bibinfo {author}
  {\bibfnamefont {G.}~\bibnamefont {Matten}}, \bibinfo {author} {\bibfnamefont
  {M.}~\bibnamefont {Ouisse}},\ and\ \bibinfo {author} {\bibfnamefont
  {M.}~\bibnamefont {Versaevel}},\ }\bibfield  {title} {\bibinfo {title} {{Duct
  modes damping through an adjustable electroacoustic liner under grazing
  incidence}},\ }\href {https://doi.org/10.1016/j.jsv.2018.04.009} {\bibfield
  {journal} {\bibinfo  {journal} {Journal of Sound and Vibration}\ }\textbf
  {\bibinfo {volume} {426}},\ \bibinfo {pages} {19} (\bibinfo {year}
  {2018})}\BibitemShut {NoStop}%
\bibitem [{\citenamefont {Rivet}\ \emph {et~al.}(2018)\citenamefont {Rivet},
  \citenamefont {Brandst{\"{o}}tter}, \citenamefont {Makris}, \citenamefont
  {Lissek}, \citenamefont {Rotter},\ and\ \citenamefont {Fleury}}]{Rivet2018a}%
  \BibitemOpen
  \bibfield  {author} {\bibinfo {author} {\bibfnamefont {E.}~\bibnamefont
  {Rivet}}, \bibinfo {author} {\bibfnamefont {A.}~\bibnamefont
  {Brandst{\"{o}}tter}}, \bibinfo {author} {\bibfnamefont {K.~G.}\ \bibnamefont
  {Makris}}, \bibinfo {author} {\bibfnamefont {H.}~\bibnamefont {Lissek}},
  \bibinfo {author} {\bibfnamefont {S.}~\bibnamefont {Rotter}},\ and\ \bibinfo
  {author} {\bibfnamefont {R.}~\bibnamefont {Fleury}},\ }\bibfield  {title}
  {\bibinfo {title} {{Constant-pressure sound waves in non-Hermitian disordered
  media}},\ }\href {https://doi.org/10.1038/s41567-018-0188-7} {\bibfield
  {journal} {\bibinfo  {journal} {Nature Physics}\ }\textbf {\bibinfo {volume}
  {14}},\ \bibinfo {pages} {942} (\bibinfo {year} {2018})}\BibitemShut
  {NoStop}%
\bibitem [{\citenamefont {Lissek}\ \emph {et~al.}(2018)\citenamefont {Lissek},
  \citenamefont {Rivet}, \citenamefont {Laurence},\ and\ \citenamefont
  {Fleury}}]{Lissek2018}%
  \BibitemOpen
  \bibfield  {author} {\bibinfo {author} {\bibfnamefont {H.}~\bibnamefont
  {Lissek}}, \bibinfo {author} {\bibfnamefont {E.}~\bibnamefont {Rivet}},
  \bibinfo {author} {\bibfnamefont {T.}~\bibnamefont {Laurence}},\ and\
  \bibinfo {author} {\bibfnamefont {R.}~\bibnamefont {Fleury}},\ }\bibfield
  {title} {\bibinfo {title} {{Toward wideband steerable acoustic metasurfaces
  with arrays of active electroacoustic resonators}},\ }\bibfield  {journal}
  {\bibinfo  {journal} {Journal of Applied Physics}\ }\textbf {\bibinfo
  {volume} {123}},\ \href {https://doi.org/10.1063/1.5011380}
  {10.1063/1.5011380} (\bibinfo {year} {2018})\BibitemShut {NoStop}%
\bibitem [{\citenamefont {Su}\ \emph {et~al.}(1979)\citenamefont {Su},
  \citenamefont {Schrieffer},\ and\ \citenamefont {Heeger}}]{Su1979}%
  \BibitemOpen
  \bibfield  {author} {\bibinfo {author} {\bibfnamefont {W.~P.}\ \bibnamefont
  {Su}}, \bibinfo {author} {\bibfnamefont {J.~R.}\ \bibnamefont {Schrieffer}},\
  and\ \bibinfo {author} {\bibfnamefont {A.~J.}\ \bibnamefont {Heeger}},\
  }\bibfield  {title} {\bibinfo {title} {{Solitons in polyacetylene}},\ }\href
  {https://doi.org/10.1103/PhysRevLett.42.1698} {\bibfield  {journal} {\bibinfo
   {journal} {Physical Review Letters}\ }\textbf {\bibinfo {volume} {42}},\
  \bibinfo {pages} {1698} (\bibinfo {year} {1979})}\BibitemShut {NoStop}%
\bibitem [{\citenamefont {Coutant}\ \emph {et~al.}(2021)\citenamefont
  {Coutant}, \citenamefont {Sivadon}, \citenamefont {Zheng}, \citenamefont
  {Achilleos}, \citenamefont {Richoux}, \citenamefont {Theocharis},\ and\
  \citenamefont {Pagneux}}]{Coutant2021a}%
  \BibitemOpen
  \bibfield  {author} {\bibinfo {author} {\bibfnamefont {A.}~\bibnamefont
  {Coutant}}, \bibinfo {author} {\bibfnamefont {A.}~\bibnamefont {Sivadon}},
  \bibinfo {author} {\bibfnamefont {L.}~\bibnamefont {Zheng}}, \bibinfo
  {author} {\bibfnamefont {V.}~\bibnamefont {Achilleos}}, \bibinfo {author}
  {\bibfnamefont {O.}~\bibnamefont {Richoux}}, \bibinfo {author} {\bibfnamefont
  {G.}~\bibnamefont {Theocharis}},\ and\ \bibinfo {author} {\bibfnamefont
  {V.}~\bibnamefont {Pagneux}},\ }\bibfield  {title} {\bibinfo {title}
  {{Acoustic Su-Schrieffer-Heeger lattice: Direct mapping of acoustic
  waveguides to the Su-Schrieffer-Heeger model}},\ }\href
  {https://doi.org/10.1103/PhysRevB.103.224309} {\bibfield  {journal} {\bibinfo
   {journal} {Physical Review B}\ }\textbf {\bibinfo {volume} {103}},\ \bibinfo
  {pages} {1} (\bibinfo {year} {2021})},\ \Eprint
  {https://arxiv.org/abs/2103.03859} {arXiv:2103.03859} \BibitemShut {NoStop}%
\bibitem [{\citenamefont {Zheng}\ \emph {et~al.}(2019)\citenamefont {Zheng},
  \citenamefont {Achilleos}, \citenamefont {Richoux}, \citenamefont
  {Theocharis},\ and\ \citenamefont {Pagneux}}]{Zheng2019}%
  \BibitemOpen
  \bibfield  {author} {\bibinfo {author} {\bibfnamefont {L.~Y.}\ \bibnamefont
  {Zheng}}, \bibinfo {author} {\bibfnamefont {V.}~\bibnamefont {Achilleos}},
  \bibinfo {author} {\bibfnamefont {O.}~\bibnamefont {Richoux}}, \bibinfo
  {author} {\bibfnamefont {G.}~\bibnamefont {Theocharis}},\ and\ \bibinfo
  {author} {\bibfnamefont {V.}~\bibnamefont {Pagneux}},\ }\bibfield  {title}
  {\bibinfo {title} {{Observation of Edge Waves in a Two-Dimensional
  Su-Schrieffer-Heeger Acoustic Network}},\ }\href
  {https://doi.org/10.1103/PHYSREVAPPLIED.12.034014/FIGURES/4/MEDIUM}
  {\bibfield  {journal} {\bibinfo  {journal} {Physical Review Applied}\
  }\textbf {\bibinfo {volume} {12}},\ \bibinfo {pages} {034014} (\bibinfo
  {year} {2019})},\ \Eprint {https://arxiv.org/abs/1903.11961}
  {arXiv:1903.11961} \BibitemShut {NoStop}%
\bibitem [{\citenamefont {Li}\ \emph {et~al.}(2018)\citenamefont {Li},
  \citenamefont {Meng}, \citenamefont {Wu}, \citenamefont {Yan}, \citenamefont
  {Huang}, \citenamefont {Wang},\ and\ \citenamefont {Wen}}]{Li2018}%
  \BibitemOpen
  \bibfield  {author} {\bibinfo {author} {\bibfnamefont {X.}~\bibnamefont
  {Li}}, \bibinfo {author} {\bibfnamefont {Y.}~\bibnamefont {Meng}}, \bibinfo
  {author} {\bibfnamefont {X.}~\bibnamefont {Wu}}, \bibinfo {author}
  {\bibfnamefont {S.}~\bibnamefont {Yan}}, \bibinfo {author} {\bibfnamefont
  {Y.}~\bibnamefont {Huang}}, \bibinfo {author} {\bibfnamefont
  {S.}~\bibnamefont {Wang}},\ and\ \bibinfo {author} {\bibfnamefont
  {W.}~\bibnamefont {Wen}},\ }\bibfield  {title} {\bibinfo {title}
  {{Su-Schrieffer-Heeger model inspired acoustic interface states and edge
  states}},\ }\href {https://doi.org/10.1063/1.5051523} {\bibfield  {journal}
  {\bibinfo  {journal} {Applied Physics Letters}\ }\textbf {\bibinfo {volume}
  {113}},\ \bibinfo {pages} {203501} (\bibinfo {year} {2018})}\BibitemShut
  {NoStop}%
\bibitem [{\citenamefont {Richoux}\ and\ \citenamefont
  {Pagneux}(2002)}]{Richoux2002}%
  \BibitemOpen
  \bibfield  {author} {\bibinfo {author} {\bibfnamefont {O.}~\bibnamefont
  {Richoux}}\ and\ \bibinfo {author} {\bibfnamefont {V.}~\bibnamefont
  {Pagneux}},\ }\bibfield  {title} {\bibinfo {title} {{Acoustic
  characterization of the Hofstadter butterfly with resonant scatterers}},\
  }\href@noop {} {\bibfield  {journal} {\bibinfo  {journal} {Europhys. Lett}\
  }\textbf {\bibinfo {volume} {59}},\ \bibinfo {pages} {34} (\bibinfo {year}
  {2002})}\BibitemShut {NoStop}%
\bibitem [{\citenamefont {Yves}\ \emph {et~al.}(2017)\citenamefont {Yves},
  \citenamefont {Fleury}, \citenamefont {Lemoult}, \citenamefont {Fink},\ and\
  \citenamefont {Lerosey}}]{Yves2017}%
  \BibitemOpen
  \bibfield  {author} {\bibinfo {author} {\bibfnamefont {S.}~\bibnamefont
  {Yves}}, \bibinfo {author} {\bibfnamefont {R.}~\bibnamefont {Fleury}},
  \bibinfo {author} {\bibfnamefont {F.}~\bibnamefont {Lemoult}}, \bibinfo
  {author} {\bibfnamefont {M.}~\bibnamefont {Fink}},\ and\ \bibinfo {author}
  {\bibfnamefont {G.}~\bibnamefont {Lerosey}},\ }\bibfield  {title} {\bibinfo
  {title} {{Topological acoustic polaritons: Robust sound manipulation at the
  subwavelength scale}},\ }\bibfield  {journal} {\bibinfo  {journal} {New
  Journal of Physics}\ }\textbf {\bibinfo {volume} {19}},\ \href
  {https://doi.org/10.1088/1367-2630/aa66f8} {10.1088/1367-2630/aa66f8}
  (\bibinfo {year} {2017})\BibitemShut {NoStop}%
\bibitem [{\citenamefont {Guo}\ \emph {et~al.}(2020)\citenamefont {Guo},
  \citenamefont {Lissek},\ and\ \citenamefont {Fleury}}]{Guo2020}%
  \BibitemOpen
  \bibfield  {author} {\bibinfo {author} {\bibfnamefont {X.}~\bibnamefont
  {Guo}}, \bibinfo {author} {\bibfnamefont {H.}~\bibnamefont {Lissek}},\ and\
  \bibinfo {author} {\bibfnamefont {R.}~\bibnamefont {Fleury}},\ }\bibfield
  {title} {\bibinfo {title} {{Improving Sound Absorption Through Nonlinear
  Active Electroacoustic Resonators}},\ }\href
  {https://doi.org/10.1103/PhysRevApplied.13.014018} {\bibfield  {journal}
  {\bibinfo  {journal} {Physical Review Applied}\ }\textbf {\bibinfo {volume}
  {13}},\ \bibinfo {pages} {1} (\bibinfo {year} {2020})},\ \Eprint
  {https://arxiv.org/abs/1907.07474} {arXiv:1907.07474} \BibitemShut {NoStop}%
\bibitem [{\citenamefont {Standards}(2011)}]{Standards2011}%
  \BibitemOpen
  \bibfield  {author} {\bibinfo {author} {\bibfnamefont {A.}~\bibnamefont
  {Standards}},\ }\bibfield  {title} {\bibinfo {title} {{Standard Test Method
  for Measurement of Normal Incidence Sound Transmission of Acoustical
  Materials Based on the Transfer Matrix Method 1}},\ }\href
  {https://webstore.ansi.org/Standards/ASTM/astme261109
  http://infostore.saiglobal.com/store/details.aspx?ProductID=1115068}
  {\bibfield  {journal} {\bibinfo  {journal} {Annual Book of ASTM Standards}\
  }\textbf {\bibinfo {volume} {i}},\ \bibinfo {pages} {1} (\bibinfo {year}
  {2011})}\BibitemShut {NoStop}%
\bibitem [{\citenamefont {Guo}\ \emph {et~al.}(2022)\citenamefont {Guo},
  \citenamefont {Volery},\ and\ \citenamefont {Lissek}}]{Guo2022}%
  \BibitemOpen
  \bibfield  {author} {\bibinfo {author} {\bibfnamefont {X.}~\bibnamefont
  {Guo}}, \bibinfo {author} {\bibfnamefont {M.}~\bibnamefont {Volery}},\ and\
  \bibinfo {author} {\bibfnamefont {H.}~\bibnamefont {Lissek}},\ }\bibfield
  {title} {\bibinfo {title} {{PID-like active impedance control for
  electroacoustic resonators to design tunable single-degree-of-freedom sound
  absorbers}},\ }\href {https://doi.org/10.1016/j.jsv.2022.116784} {\bibfield
  {journal} {\bibinfo  {journal} {Journal of Sound and Vibration}\ }\textbf
  {\bibinfo {volume} {525}},\ \bibinfo {pages} {116784} (\bibinfo {year}
  {2022})}\BibitemShut {NoStop}%
\bibitem [{\citenamefont {Maxime}\ \emph {et~al.}(2023)\citenamefont {Maxime},
  \citenamefont {Xinxin},\ and\ \citenamefont {Herv{\'{e}}}}]{Volery2023}%
  \BibitemOpen
  \bibfield  {author} {\bibinfo {author} {\bibfnamefont {V.}~\bibnamefont
  {Maxime}}, \bibinfo {author} {\bibfnamefont {G.}~\bibnamefont {Xinxin}},\
  and\ \bibinfo {author} {\bibfnamefont {L.}~\bibnamefont {Herv{\'{e}}}},\
  }\bibfield  {title} {\bibinfo {title} {{Robust direct acoustic impedance
  control using two microphones for mixed feedforward-feedback controller}},\
  }\href {https://doi.org/10.1051/aacus/2022058} {\bibfield  {journal}
  {\bibinfo  {journal} {Acta Acust.}\ }\textbf {\bibinfo {volume} {7}},\
  \bibinfo {pages} {2} (\bibinfo {year} {2023})}\BibitemShut {NoStop}%
\bibitem [{\citenamefont {Richoux}\ \emph {et~al.}(2009)\citenamefont
  {Richoux}, \citenamefont {Morand},\ and\ \citenamefont
  {Simon}}]{Richoux2009}%
  \BibitemOpen
  \bibfield  {author} {\bibinfo {author} {\bibfnamefont {O.}~\bibnamefont
  {Richoux}}, \bibinfo {author} {\bibfnamefont {E.}~\bibnamefont {Morand}},\
  and\ \bibinfo {author} {\bibfnamefont {L.}~\bibnamefont {Simon}},\ }\bibfield
   {title} {\bibinfo {title} {{Analytical study of the propagation of acoustic
  waves in a 1D weakly disordered lattice}},\ }\href
  {https://doi.org/10.1016/j.aop.2009.05.011} {\bibfield  {journal} {\bibinfo
  {journal} {Annals of Physics}\ }\textbf {\bibinfo {volume} {324}},\ \bibinfo
  {pages} {1983} (\bibinfo {year} {2009})}\BibitemShut {NoStop}%
\bibitem [{\citenamefont {Wang}\ and\ \citenamefont {Mak}(2012)}]{Wang2012}%
  \BibitemOpen
  \bibfield  {author} {\bibinfo {author} {\bibfnamefont {X.}~\bibnamefont
  {Wang}}\ and\ \bibinfo {author} {\bibfnamefont {C.~M.}\ \bibnamefont {Mak}},\
  }\bibfield  {title} {\bibinfo {title} {{Acoustic performance of a duct loaded
  with identical resonators}},\ }\href {https://doi.org/10.1121/1.3691826}
  {\bibfield  {journal} {\bibinfo  {journal} {The Journal of the Acoustical
  Society of America}\ }\textbf {\bibinfo {volume} {131}},\ \bibinfo {pages}
  {EL316} (\bibinfo {year} {2012})}\BibitemShut {NoStop}%
\bibitem [{\citenamefont {Jim{\'{e}}nez}\ \emph {et~al.}(2021)\citenamefont
  {Jim{\'{e}}nez}, \citenamefont {Groby},\ and\ \citenamefont
  {Romero-Garc{\'{i}}a}}]{Jimenez2021}%
  \BibitemOpen
  \bibfield  {author} {\bibinfo {author} {\bibfnamefont {N.}~\bibnamefont
  {Jim{\'{e}}nez}}, \bibinfo {author} {\bibfnamefont {J.~P.}\ \bibnamefont
  {Groby}},\ and\ \bibinfo {author} {\bibfnamefont {V.}~\bibnamefont
  {Romero-Garc{\'{i}}a}},\ }\bibfield  {title} {\bibinfo {title} {{The Transfer
  Matrix Method in Acoustics: Modelling One-Dimensional Acoustic Systems,
  Phononic Crystals and Acoustic Metamaterials}},\ }\href
  {https://doi.org/10.1007/978-3-030-84300-7_4} {\bibfield  {journal} {\bibinfo
   {journal} {Topics in Applied Physics}\ }\textbf {\bibinfo {volume} {143}},\
  \bibinfo {pages} {103} (\bibinfo {year} {2021})}\BibitemShut {NoStop}%
\bibitem [{\citenamefont {{De Bono}}\ \emph {et~al.}(2022)\citenamefont {{De
  Bono}}, \citenamefont {Collet}, \citenamefont {Matten}, \citenamefont
  {Karkar}, \citenamefont {Lissek}, \citenamefont {Ouisse}, \citenamefont
  {Billon}, \citenamefont {Laurence},\ and\ \citenamefont
  {Volery}}]{DeBono2022}%
  \BibitemOpen
  \bibfield  {author} {\bibinfo {author} {\bibfnamefont {E.}~\bibnamefont {{De
  Bono}}}, \bibinfo {author} {\bibfnamefont {M.}~\bibnamefont {Collet}},
  \bibinfo {author} {\bibfnamefont {G.}~\bibnamefont {Matten}}, \bibinfo
  {author} {\bibfnamefont {S.}~\bibnamefont {Karkar}}, \bibinfo {author}
  {\bibfnamefont {H.}~\bibnamefont {Lissek}}, \bibinfo {author} {\bibfnamefont
  {M.}~\bibnamefont {Ouisse}}, \bibinfo {author} {\bibfnamefont
  {K.}~\bibnamefont {Billon}}, \bibinfo {author} {\bibfnamefont
  {T.}~\bibnamefont {Laurence}},\ and\ \bibinfo {author} {\bibfnamefont
  {M.}~\bibnamefont {Volery}},\ }\bibfield  {title} {\bibinfo {title} {{Effect
  of time delay on the impedance control of a pressure-based, current-driven
  Electroacoustic Absorber}},\ }\href
  {https://doi.org/10.1016/j.jsv.2022.117201} {\bibfield  {journal} {\bibinfo
  {journal} {Journal of Sound and Vibration}\ }\textbf {\bibinfo {volume}
  {537}},\ \bibinfo {pages} {117201} (\bibinfo {year} {2022})}\BibitemShut
  {NoStop}%
\end{thebibliography}%

\end{document}


\maketitle
\beginsupplement

\section{Finite element simulations}
\begin{figure}[ht]
	\centering
	\begin{subfigure}[h]{0.33\textwidth}
		\centering
		\includegraphics[width=\textwidth]{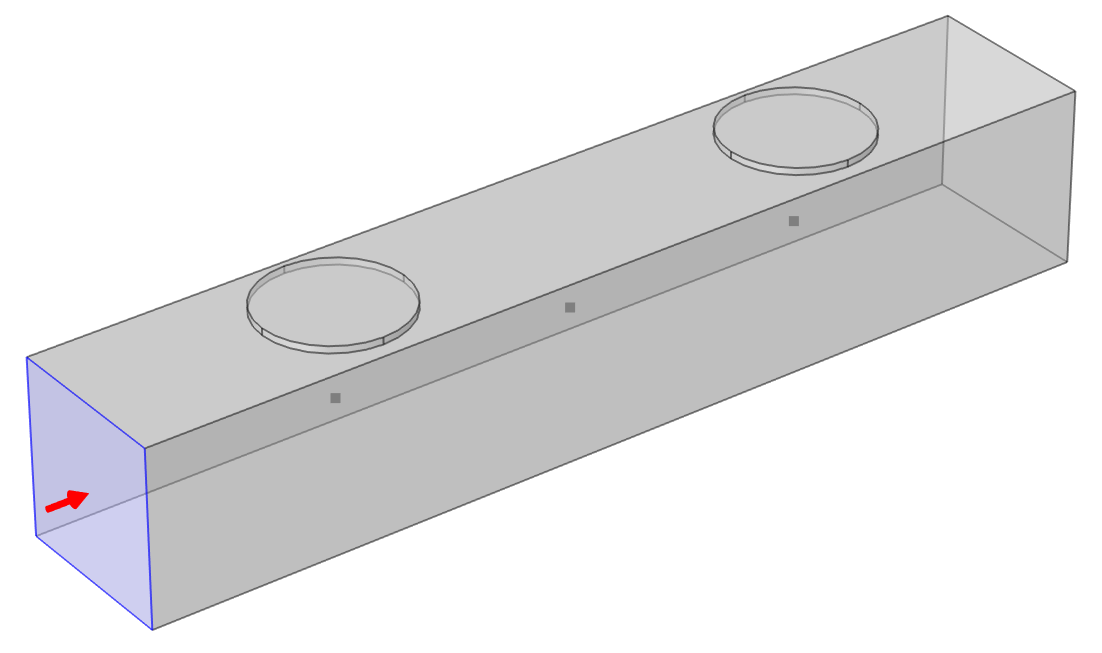}
		\caption{Unitcell model}
	\end{subfigure}\\
	\begin{subfigure}[h]{0.66\textwidth}
		\centering
		\includegraphics[width =\textwidth]{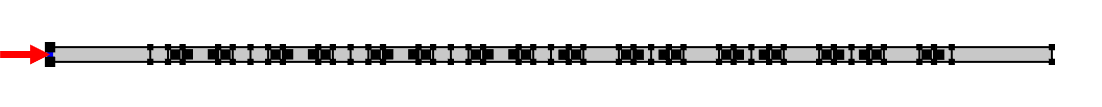}
		\caption{Meta-crystal model}
	\end{subfigure}%
	\caption{\textcolor{blue}{Geometries used for the COMSOL finite element simulations. The red arrows show to the inlet port (- the outlet is on the opposing extremity).} \label{fig:COMSOL}}
\end{figure}

 COMSOL multiphysics finite element simulations were performed enabling design optimization and experimental validation. \textcolor{blue}{Using the "Pressure Acoustics, Frequency Domain" module, each lined speaker is modelled using experimentally obtained lumped parameters (c.f. Section \ref{sec:Z_mech}). Furthermore, the electronic control feedback delay is taken into account to faithfully capture the physics. The scattering matrix elements S11 and S12 (- R and T respectively) are obtained using ports at the extremities of the simulated unitcell and meta-crystal (- Fig.~\ref{fig:COMSOL}).}

\textcolor{blue}{From the unitcell finite element simulation, the dispersion could be computed and compared to the analytical predictions and the finite element simulation and the experimental measurement of the are plotted in the following figures:}
\begin{figure}[ht]
	\centering
	\begin{subfigure}[h]{0.33\textwidth}
		\centering
		\includegraphics[width=\textwidth]{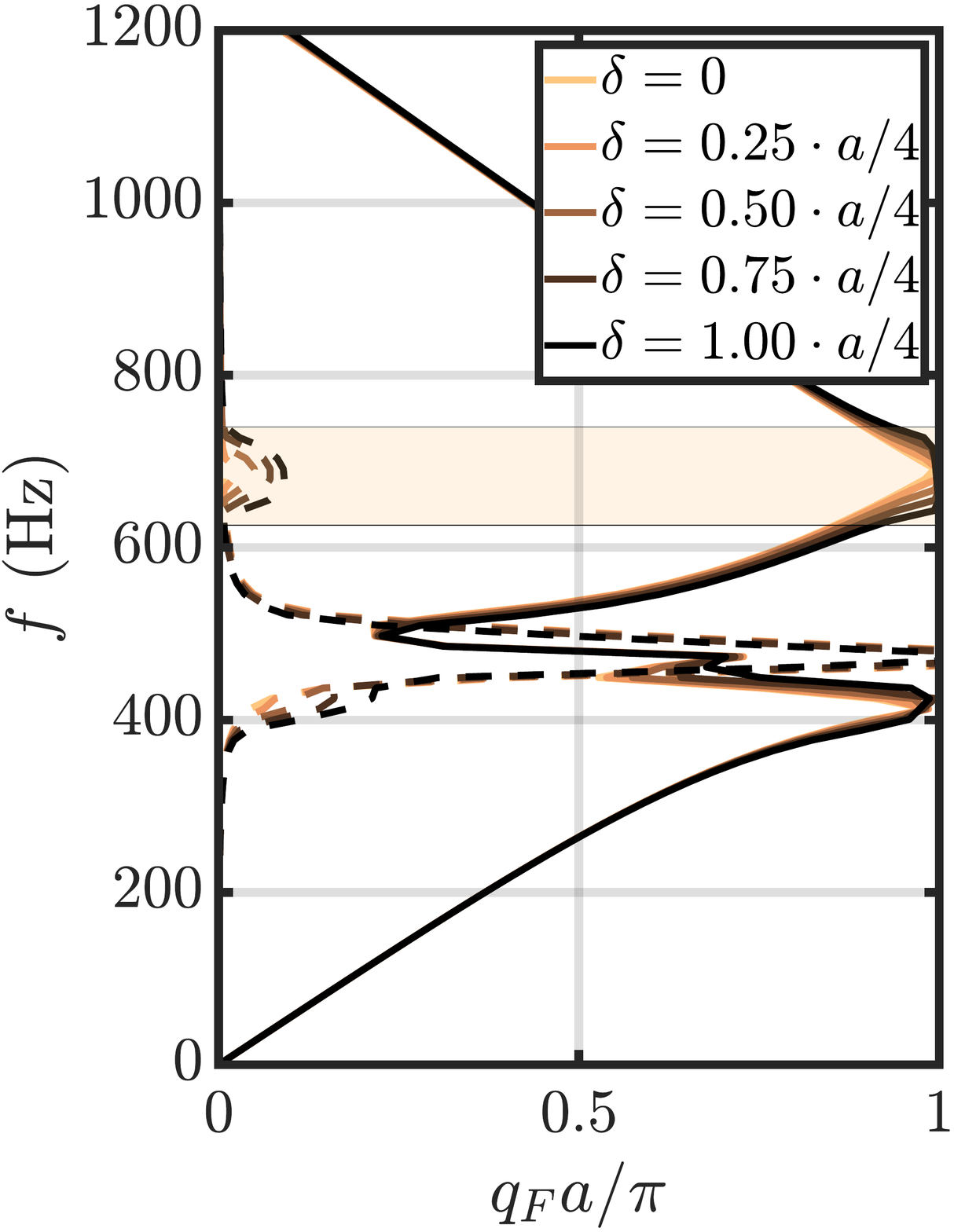}
		\caption{Analytical sim.}
		\label{fig::results_ana}
	\end{subfigure}%
	~ 
	\begin{subfigure}[h]{0.33\textwidth}
		\centering
		\includegraphics[width =\textwidth]{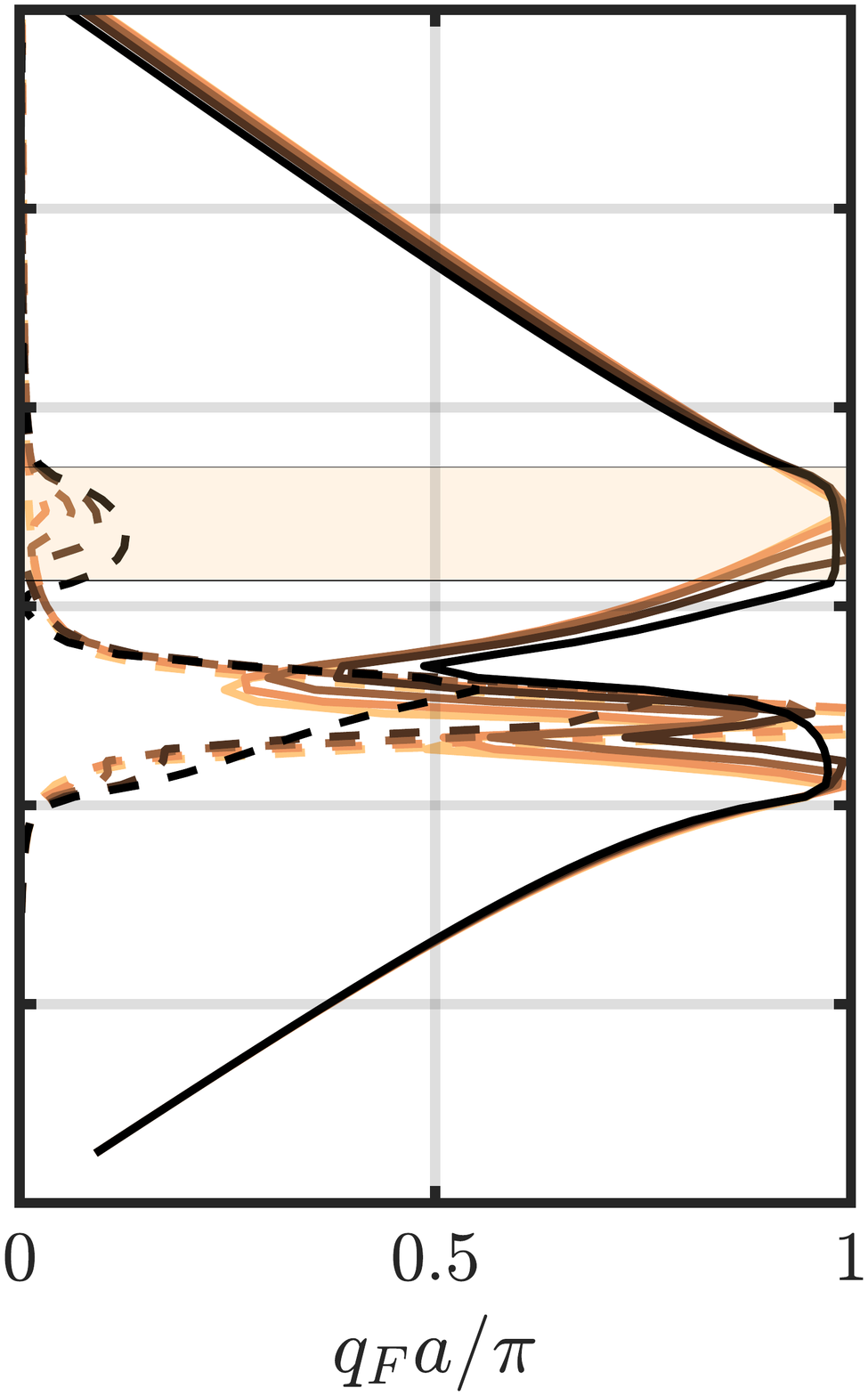}
		\caption{Finite element sim.}
		\label{fig::results_sim}
	\end{subfigure}%
	~ 
	\begin{subfigure}[h]{0.33\textwidth}
		\centering
		\includegraphics[width =\textwidth]{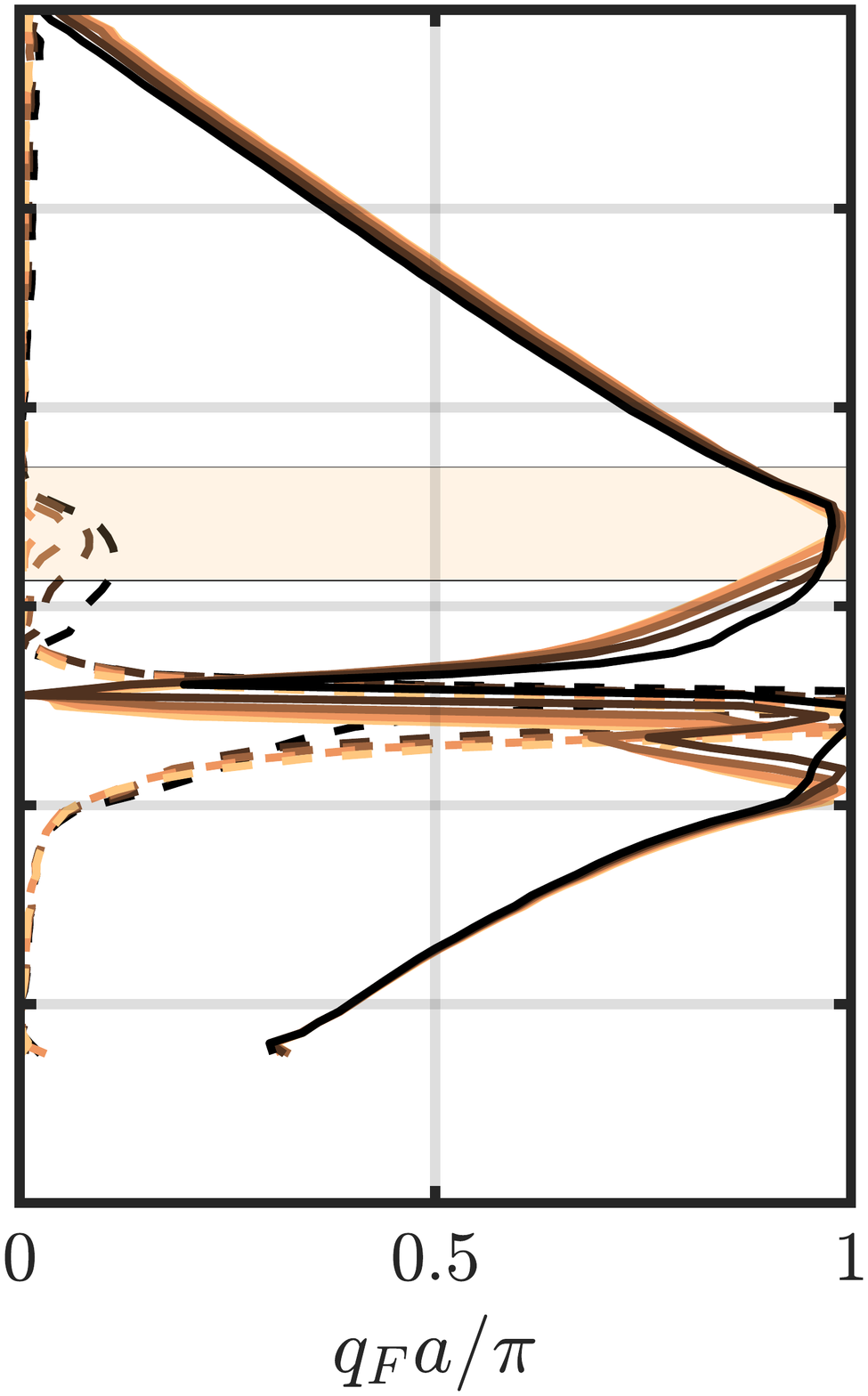}
		\caption{Experimental}
		\label{fig:results_exp}
	\end{subfigure}
	\caption{Active control on the band gap size. The dispersion relation of the acoustic ssh for five values of the virtual shift $\tilde{\delta}$ derived from  (a) the analytical model, (b) the finite element model (comsol multiphysics) and (c) measured experimental data. solid line and dashed line represent the real and imaginary parts of the bloch-floquet wave number $q$ respectively. The frequency $f$ which is related to the wavenumber $k$ through $k = \frac{2\pi}{c} f$.}\label{fig::disp}
\end{figure}

\newpage

\section{Experimental transfer matrix $M_{cell}$ measurement}

\begin{figure}[ht]
	\centering
	\includegraphics[width =  0.5\textwidth]{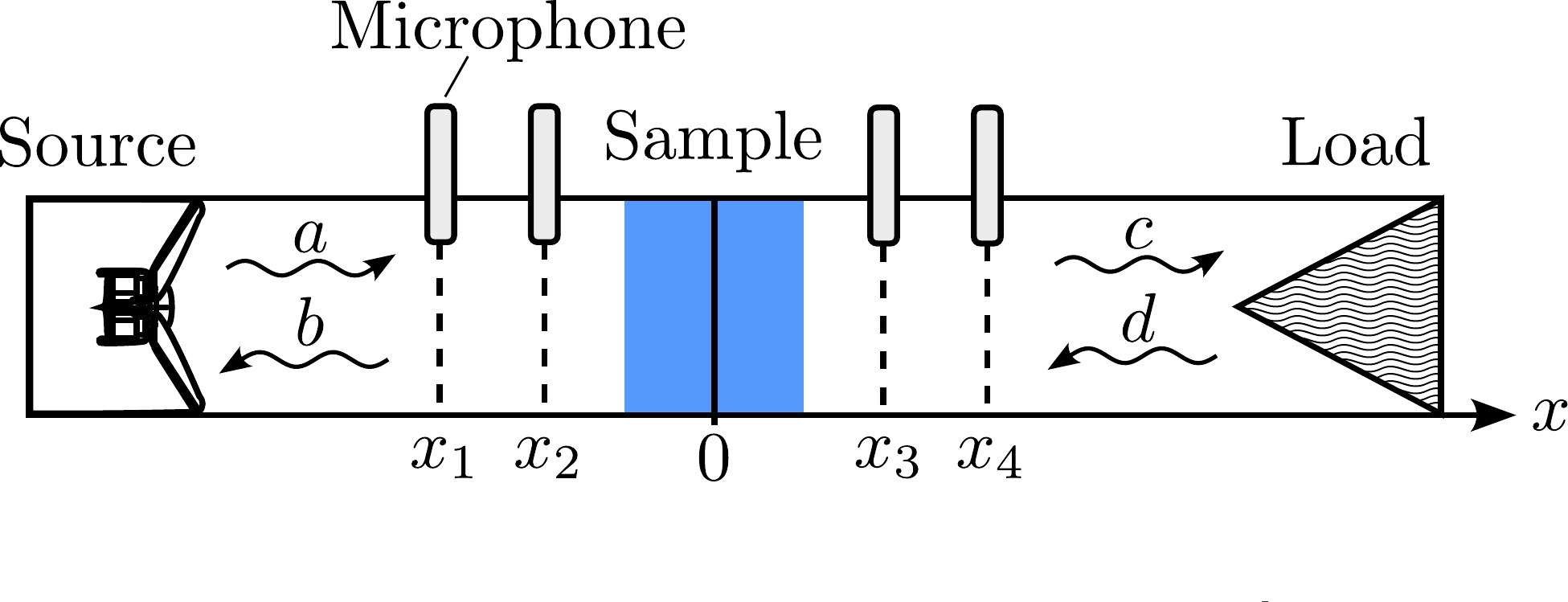}
	\caption{Characterisation of the transfer matrix of a sample using a loaded Kundt duct and four pressure microphones. Spatial separation of the microphones on both sides of the sample allows for the determination of the forward (a,c) and backward (b,d) complex field amplitudes.}
\end{figure}
Complex pressures $P_i$ measured at $x_1,x_2,x_3 \mbox{ and } x_4$:
\begin{equation}
	\begin{aligned}
		&P_{1}=\left(a e^{-j k x_{1}}+b e^{j k x_{1}}\right) e^{j \omega t} \\
		&P_{2}=\left(a e^{-j k x_{2}}+b e^{j k x_{2}}\right) e^{j \omega t} \\
		&P_{3}=\left(c e^{-j k x_{3}}+d e^{j k x_{3}}\right) e^{j \omega t} \\
		&P_{4}=\left(d e^{-j k x_{4}}+d e^{j k x_{4}}\right) e^{j \omega t}
	\end{aligned}
\end{equation}

where the coefficients can be extracted:
\begin{equation}
		\begin{aligned}
		&a=\frac{j\left(P_{1} e^{j k x_{2}}-P_{2} e^{j k x_{1}}\right)}{2 \sin(k |x_{1}-x_{2}|)} \\
		&b=\frac{j\left(P_{2} e^{-j k x_{1}}-P_{1} e^{-j k x_{2}}\right)}{2 \sin(k|x_{1}-x_{2}|)} \\
		&c=\frac{j\left(P_{3} e^{j k x_{4}}-P_{4} e^{j k x_{3}}\right)}{2 \sin(k|x_{3}-x_{4}|)} \\
		&d=\frac{j\left(P_{4} e^{-j k x_{3}}-P_{3} e^{-j k x_{4}}\right)}{2 \sin(k|x_{3}-x_{4}|)}
	\end{aligned}
\end{equation}

The scattering matrix links the outgoing complex field amplitude to the ingoing:

\begin{equation}
	\left(\begin{array}{c}
		be^{jka/2} \\
		ce^{-jka/2} 
	\end{array}\right)_{\text {Outgoing }}=\left(\begin{array}{cc}
		S_{11} & S_{12} \\
		S_{21} & S_{22}
	\end{array}\right)\left(\begin{array}{c}
		ae^{jka/2} \\
		de^{-jka/2}
	\end{array}\right)_{\text {Ingoing}}
\end{equation}

If reciprocal ($\rightarrow S_{12}=S_{21}$) and symmetrical ($\rightarrow S_{11}=S_{22}$):
\begin{equation}
	\begin{gathered}
		S_{11}=S_{22}=\frac{a b e^{jka}-c d e^{-jka}}{a^{2}e^{jka}-d^{2}e^{-jka}} \\
		S_{12}=S_{21}=\frac{a c-b d }{a^{2}e^{jka}-d^{2}e^{-jka}}
	\end{gathered}
\end{equation}

If the system is neither reciprocal nor symmetrical, another set of measurements is required in order to resolve all four elements of the scattering matrix.


Finally, the transfer matrix $M_{cell}$ is directly obtained from $S$:

\begin{equation}
	M_{cell} =\left(\begin{array}{ll}
		S_{21}-\frac{S_{22} S_{11}}{S_{12}} & \frac{S_{22}}{S_{12}} \\
		\frac{-S_{11}}{S_{12}} & \frac{1}{S_{12}}
	\end{array}\right)
\end{equation}

\color{blue}
	
\section{Mechanical parameter characterization\label{sec:Z_mech}}

Characterization of the mechanical parameters of a closed-box electrodynamic loudspeaker was achieved by direct measurement of the impedance following methods described by E. Rivet \cite{Rivet2017a} using quarter-inch PCB microphones and a single-point Polytec laser vibrometer for simultaneous pressure and velocity measurement respectively. A Tannoy source speaker was used to generate sound pressure $p_f$ at the diaphragm resulting in electrical current $i$ at its electrical terminals (- in a closed circuit configuration). Assuming steady-state and expressed in the frequency domain, its velocity response $v(s)$ is derived from Newton's second law of motion:

\begin{equation}
	\zeta_{mc}(s)v(s) = S_d p_f(s) - B\ell i(s)
\end{equation}

Where:
\begin{itemize}
	\item $\zeta_{mc}$: total mechanical impedance (N.s/m), including the compliance of air inside the cabinet
	\item $v$: diaphragm velocity (m/s)
	\item $S_d$: diaphragm surface area ($m^2$)
	\item $p_f$: pressure in front of diaphragm (Pa)
	\item $B\ell$: Force factor (T.m)
	\item $i$: electrical current (A)
	\item $s = j \omega$: the Laplace variable (Rad/s)
\end{itemize}

Using lumped parameters to model the speaker, the mechanical impedance is written as:

	\begin{equation}
	\zeta_{mc}(s) = M_{ms}\cdot s + R_{ms} + 1/(C_{mc}\cdot s)
\end{equation}
	Where:
\begin{itemize}
	\item $M_{ms}$: diaphragm mass (kg)
	\item $R_{ms}$: mechanical resistance  
	\item $C_{mc}$: speaker + cabinet compliance ($s^2$/kg)
\end{itemize}

\begin{figure}[ht]
	\centering
	\includegraphics[width =  0.5\textwidth]{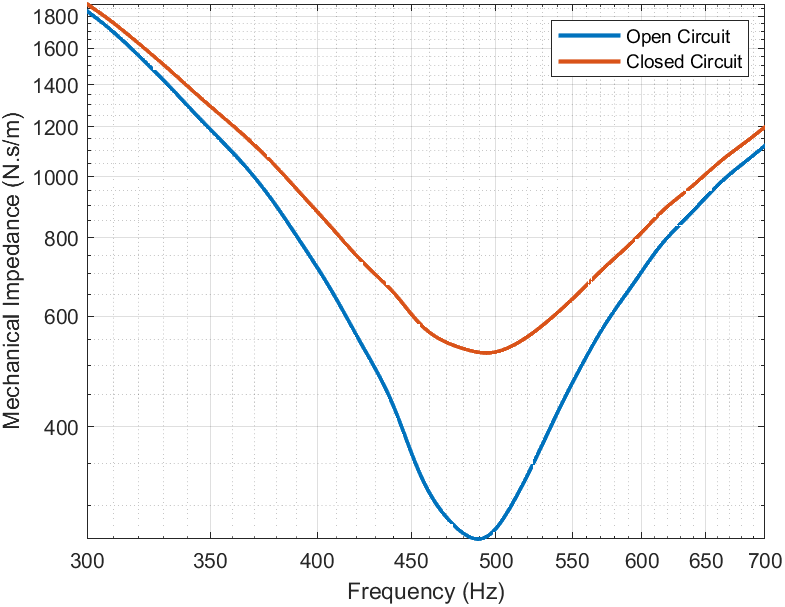}
	\caption{\textcolor{blue}{Measured impedance for both open end closed circuit configurations. $Bl = 1.468620e+00$; $R_{ms} = 3.685479e-01$, $M_{ms} = 5.941674e-04$; $C_{mc} = 1.707260e-04$\label{fig:Z_mech}}}
\end{figure}

Knowing $S_d$, the mechanical parameters and the force factor can be estimated by fitting and subtracting the measured impedance of both open ($i=0$) and closed circuit ($i\neq0$) configurations as shown in figure~\ref{fig:Z_mech}.

\bibliography{library}
